\documentclass[12pt]{article}

\usepackage{scicite}

\usepackage{times}
\usepackage{xcolor}
\usepackage{graphicx}
\usepackage[hyphens]{url}
\usepackage{hyperref}
\hypersetup{breaklinks=true}
\usepackage{float}
\usepackage[skip=2pt]{subcaption}
\usepackage{xspace}
\usepackage{enumitem}
\usepackage{amsmath}

\usepackage{tabularx}
\usepackage{booktabs}

\newcolumntype{Y}{>{\centering\arraybackslash}X}
\newcolumntype{M}[1]{>{\centering\arraybackslash}m{#1}}
\allowdisplaybreaks[1]

\usepackage{placeins}
\usepackage{fancyhdr}

\fancypagestyle{firststyle}
{
    \fancyfoot{}
    \fancyfoot[R]{\thepage}
    \fancyfoot[L]{COVID-19 School Reopening Strategies}
}

\newenvironment{sciabstract}{%
\begin{quote} \bf}
{\end{quote}}

\title{The impact of school reopening strategies during COVID-19: A case study of São Paulo, Brazil}

\author
{E. H. M. CRUZ $^{1\ast}$, J. M. MACIEL$^{1}$, C. CLOZATO$^{1}$, \\ M. S. SERPA$^{2}$, P. O. A. NAVAUX$^{2}$, E. MENESES$^{3,4}$,\\ M. ABDALAH$^{3}$, M. DIENER$^{5}$ \\
\\
\normalsize{$^{1}$Campus Paranavaí, Federal Institute of Parana (IFPR), Paranavaí, Brazil}\\

\normalsize{$^{2}$Informatics Institute, Federal University of Rio Grande do Sul (UFRGS), Porto Alegre, Brazil}\\

\normalsize{$^{3}$Costa Rica National High Technology Center, Costa Rica}\\

\normalsize{$^{4}$Costa Rica Institute of Technology, Costa Rica}\\

\normalsize{$^{5}$University of Illinois at Urbana-Champaign, USA}\\
\\
\normalsize{$^\ast$To whom correspondence should be addressed; E-mail: eduardo.cruz@ifpr.edu.br} 
\\
\normalsize{Address: IFPR Campus Paranavaí -- Rua José Felipe Tequinha, 1400 -- Jardim das Nações}
\\
\normalsize{Paranavaí-PR, Brazil -- Postal code 87703-536}
}

\date{}

\begin{document}

\nocite{}


\baselineskip24pt


\maketitle
\thispagestyle{firststyle}

\clearpage


\begin{sciabstract}

During the COVID-19 pandemic, many countries opted for strict public health measures, including closing schools. They have now started relaxing some of those restrictions.
To avoid overwhelming health systems, predictions for the number of new COVID-19 cases need to be considered when choosing a school reopening strategy.
Using computer simulation, we analyze different strategies to reopen schools in the São Paulo Metropolitan Area, including the official reopening plan.
Our results indicate that reopening schools with all students at once has a big impact on the number of new COVID-19 cases, which could cause a collapse of the health system.
On the other hand, a controlled school reopening avoids the collapse of the health system, with a maximum ICU occupancy between 48.8\% and 97.8\%, depending on how people attending school follow sanitary measures.
Postponing the schools' return date for after a vaccine becomes available can save up to \mbox{37,753} lives just in the São Paulo Metropolitan Area.

\end{sciabstract}

\clearpage

\section*{Introduction}

In December 2019, a novel coronavirus disease (COVID-19) emerged in Wuhan, China~\cite{HUI:IJID:2020}, and has since spread out to the rest of the world, evolving into a pandemic~\cite{WOROBEY:SCIENCE:2020}. Due to its high infection rate, the virus SARS-CoV-2 is not only causing mortality, but it is also stressing national health systems due to the large number of infected people that need hospitalization, also causing a profound economic impact~\cite{MILLER:NATUREMEDICINE:2020,PHUA:LANCETRESPIRATORYMEDICINE:2020,KAWOHL:LANCETPSY:2020}.
In the absence of a vaccine or an effective treatment for COVID-19, the role of public health measures mostly comprehends the so-called \emph{non-pharmaceutical interventions} (NPIs)~\cite{imperial:report9:2020}, including social distancing and isolation, reducing economic activities, enforcing home office, moving to remote classes and closing social venues where people tend to agglomerate in close proximity.
Initially, sanitary authorities opted for strict NPIs.
As the pandemic evolved, other factors emerged and started an interplay with the health crisis.
Therefore, measures were revisited and, oftentimes, relaxed.
However, a critical question is still unresolved:
What is the impact of reopening the school system for in-person classes~\cite{science:Lordan:2020,lancet:jasmina:2020}?

When to start reopening schools and the reopening policy are particularly important questions for multiple reasons: $i)$ education is a basic human necessity;
$ii)$ in-person classes enable other economic activities because parents can go to work while their children are at school~\cite{lancet:viner:2020}, besides other markets associated with schools, such as transportation, food, clothing, among others;
$iii)$ in-person classes are often hard to replace with online learning, especially in developing countries and for impoverished families, due to lower availability of broadband Internet and fast laptop computers with cameras;
$iv)$ due to the relatively long time children spend in proximity to each other in schools, the potential for exposure is much higher, requiring several workarounds to reduce the transmission rate~\cite{vermund:2020}, especially because young people tend to be asymptomatic while carrying high viral loads~\cite{Hildenwall:2020}, which increases the chances of an exponential transmission growth, thereby requiring additional quarantines to prevent a health system collapse; and
$v)$~a well-crafted policy for returning to school may help preventing new waves of the pandemic, save lives, and reduce the amount of psychological stress due to the health crisis~\cite{LANCETEDITORIALPSY:LANCETPSY:2020}.
To understand the impact of the pandemic on schools and develop policies to deal with it, mathematical and computer simulation modeling is crucial~\cite{science:Metcalf:2020}.

In this work, we set out to evaluate a range of strategies for opening the school system of the São Paulo Metropolitan Area~(SPMA) using computer simulation.
We chose this region as a case study because Brazil is one of the current epicenters of the pandemic~\cite{BARBERIA:LANCET:2020,imperial:report21:2020}, and the SPMA is the most relevant of the country due its huge population and economical importance.
The region contains 39 cities and 21.7 million inhabitants, representing about 10\% of Brazil's population~\cite{sus:datasus:tabnet}.
We analyze 3 different strategies of school reopening: (i) reopen schools with all students at once; (ii) reopen schools following a strategy based on the official plan of the São Paulo government~\cite{PlanoSP:2020}, which consists of 3 stages, carefully increasing the amount of students in each stage; and (iii) reopen schools only when a vaccine becomes available. We show results regarding the total amount of COVID-19 cases, critical cases and number of deaths.

\section*{Materials and Methods}

Most models to analyze pandemics employ compartmental models, such as SIS, SIR, SEIR or SEIHR~\cite{AbouIsmail:2020}.
However, the assumptions of homogeneous random mixing in those models can lead to wrong projections.
Also, the effects of non-pharmaceutical interventions (NPIs)~\cite{imperial:report9:2020} are difficult to implement in those models.
To overcome these issues, we developed a heterogeneous and dynamic network stochastic model~\cite{Enright:2018, Brauer:2017, Graham:2013}.
The model was implemented as a simulator written in the C++ language, which was chosen due its high performance, low overhead and powerful object-oriented features.
The network was modeled as an undirected graph $G = (V, E)$, in which the vertices $V$ represent the people and the edges $E$ the relations between people, which have an infection rate.
Given two people $v_1$ and $v_2$, such that $v_1 \in V$ and $v_2 \in V$, if we have an edge $e \in E$ such that $e = (v_1, v_2)$, the person $v_1$ can infect the person $v_2$.
To support heterogeneous infection rates, we color the edges of the graph, such that we can attribute different infection rates for each color.
The color represents the type of relation.
Although it is possible to have any kind of relations modeled, for this work we modeled the following relation types: home, community, workplace, schools and Inter-city.

Since our model supports heterogeneous networks, it can provide a better representation of the population~\cite{science:Britton:2020}.
Each person can have an arbitrary number of edges in the graph, and the relation type (color) of each edge is also arbitrary.
Therefore, as in real life, each person can have different infection rates depending on the number of relations they have, as well as the type of each relation.
Since each person has a different infection rate, each person also has a different basic reproduction rate ($R_0$). Therefore, to set the overall $R_0$, we calibrate the infection rates over the entire network such that the average $R_0$ among all people is equal to the target value.
As our model supports dynamic networks, we can change relations at any time during the simulation.
We are able to modify the infection rates, as well as add and remove edges from the graph.
This is useful to implement sanitary interventions.

We based our model in the SEIR compartmental model~\cite{AbouIsmail:2020}, but extended it to support more compartments.
Each person can be in one of the following states:

\begin{description}[noitemsep]
\item[Susceptible] The person can be infected.
\item[Infected] The person is currently infected. An infected person can have different sub-states, as we will explain later.
\item[Immune] The person is immune, either from recovering from the disease or from taking a vaccine. The person will not contract the disease anymore.
\item[Dead] The person died.
\end{description}

As for infected people, each one can be in the following sub-states:

\begin{description}[noitemsep]
\item[Incubation] The person was infected but has not developed any symptoms yet and is not contagious.

\item[Unreported] The person is contagious but either is asymptomatic or has such light symptoms that it does not receive any health care. This person, in real life, would not be part of any official statistics, hence a sub-notified case.

\item[Pre-symptomatic] The person is contagious, but did not develop any symptoms yet.

\item[Mild] The person has mild symptoms and can recover at home.

\item[Severe] The person has severe symptoms and requires health care in a hospital.

\item[Critical] The person has critical symptoms and requires health care in an Intensive Care Unit (ICU). After recovering from critical symptoms, the state is set to severe.
\end{description}

The disease spreads from infected people to susceptible people according to the infection rate of each edge in the graph, which is evaluated using a Monte Carlo based approach.
The probability of a susceptible person $p$ to get infected in a simulation cycle is given by Equations~\ref{eq:propagation:1} to~\ref{eq:propagation:4}.
Equation~\ref{eq:propagation:3} defines the number of neighbors of each vertex (person) that are infected -- $N(p)$.
In Equation~\ref{eq:propagation:4}, the function $InfectRate(p, n)$ returns the infection rate between $p$ and $n$ in the network.
\begin{gather}
G = (V, E)
\label{eq:propagation:1}\\
p \in V
\label{eq:propagation:2}\\
N(p) = \bigcup v \in V \mid \exists e \in E, e = (v, p), state(v) = infected
\label{eq:propagation:3}\\
InfectProbability(p) = 1 - \displaystyle\prod_{n \in N(p)} (1 - InfectRate(p, n))
\label{eq:propagation:4}
\end{gather}

In the simulation, we consider all the cities and population that comprehend the SPMA, including the demographic distribution.
Data used to set the simulation parameters were extracted from the literature and official available data.
Since our model is stochastic, each execution of the program generates a different network and a different disease transmission sample path.
Therefore, our results derive from 10 executions and are shown within a 95\% confidence interval following Student's t-distribution.
A comprehensive discussion on the simulation parameters, how we generate the relation network and handle sanitary interventions, as well as figures containing additional results, can be found in the supplementary material.

\section*{Results and discussion}

In the first strategy, we evaluate what could happen if all students went back to school at once, but following sanitary measures, such as wearing masks and frequently washing their hands.
Figure~\ref{fig:strategy:all:students} presents the results for this scenario.
In the left column, we show results regarding the general population of SPMA.
In the right column, we show results regarding only people that attend school, students and teachers.
Results considering the sub-population of teachers and families of people that attend school can be found in the supplementary material.
Predicting how much the infection rate in schools differs from the infection rate in the community is a difficult task. Similarly, it is nearly impossible to predict how strictly students will follow the sanitary measures.
Consequently, we evaluate two different infection rates in schools: a best case and a worst case scenario, in which the school infection rates are 2 and 4 times the community infection rate, respectively.
Also, for these results, we consider a 50\% social isolation level inside the same classroom (we consider two students as isolated when they do not have any contact with each other, and therefore have no relation between them in the network).
We show results regarding the expected number of reported COVID-19 cases, critical cases, which require intensive care units (ICU) for their treatment, and number of deaths.
In the simulations, we consider that the health system is always able to handle the demand, such that we can evaluate the peak ICU usage; thereby, no one dies from health system collapse.
In all results, day~0 corresponds to February 26th, 2020, which is the date corresponding to the first documented case in Brazil.
School reopening is set to day~224, which corresponds to October 7th, 2020, which is when the São Paulo State plans to open schools.

Opening schools without any kind of constraint on the number of students per classroom, using its total capacity, can increase the number of reported COVID-19 cases by up to \mbox{835,639}, and \mbox{40,860} more deaths occur compared to when schools do not reopen, with a peak of \mbox{12,748} people in critical condition that will require ICU treatment.
In the SPMA, there are about \mbox{5,000} ICUs available for COVID-19~\cite{seade:github}.
Therefore, the health care system would collapse in case all students returned to school at once.
By analyzing the data only in the school sub-population, we first note that the ratio of critical patients and deaths are small compared to the entire population.
While the average death rate of infected people in the population was \mbox{5.3\%}, the death rate in the school sub-population was only \mbox{0.2\%}.
This is because most students are young and, according to the official COVID-19 statistics, younger people have a smaller probability to develop the critical symptoms of the disease.
Nevertheless, we can observe that the COVID-19 reported cases among people that frequent schools increase to up to \mbox{3.6$\times$} when compared to the scenario where schools do not reopen.
This result shows that, although the opening of schools with full capacity does not present a huge risk for the students, students act as vectors of the virus, causing a big impact in the entire metropolitan area.
It is also important to note that, if we consider that people in critical state that do not receive ICU treatment dies, the amount of deaths would be much higher, since the health system would collapse if this strategy were adopted.

In the second strategy, we analyze a school opening strategy based on the plan of the S\~{a}o Paulo state government~\cite{PlanoSP:2020}.
The plan consists of 3 stages.
In the first stage, about a third of the students per class attend school each day.
In the second stage, about two thirds of the students go to school.
In the last stage, all students go back to school.
Every student and professional must obey sanitary measures such as wearing masks and frequently washing their hands.
In our evaluation, stage~1 lasts 4~weeks and stage~2 lasts 17~weeks.
The results of using this strategy are depicted in Figure~\ref{fig:strategy:sp:plan}, following the same organization of the previous strategy (Figure~\ref{fig:strategy:all:students}).
For this experiment, we consider an isolation level of 50\%.
Compared to when all students go to school, we can observe that the total number of reported infected people, in the worst case, reduces by \mbox{96,216}.
This translates to \mbox{4,437} fewer deaths, as well as reduction of the critical patients peak to \mbox{4,890} (after schools opened), which represents an ICU occupancy of only 97.8\%.
In the best case, the maximum ICU occupancy is~48.8\% after schools open.

Results only considering the school sub-population follow the same trend of the general results.
Considering the worst case, we can reduce the number of reported cases by \mbox{29,200} people and the number of deaths by 54 people compared to when all students go to school.
Regarding the deaths expected for the worst case, there would be a total of 541 deaths inside the school sub-population, in which 356 deaths correspond to teachers, representing 65.8\% of school-related deaths.
Although there are many more students than teachers, teachers are older than the students, and hence have a higher mortality rate.
If students strictly follow the sanitary measures, as considered by the best case, the difference between opening the schools with all students or with São Paulo's plan is lower, but still very significant.
Overall, results show that, although the difference in the number of cases is low compared to when all students come back at once, using an opening strategy with several phases avoids the health system collapse, which would have a big impact in the actual number of deaths.

In Figure~\ref{fig:strategy:sp:explore}, we analyze several isolation levels for the São Paulo strategy.
An isolation level of 100\% means that students inside the same classroom have no interaction between each other, while an isolation level of 0\% means that each student is able to infect any other student inside the same classroom.
As in previous experiments, we evaluate two different  rates.
If students follow the social distancing and sanitary measures very strictly, with a 100\% isolation, there would be up to \mbox{678,729} less reported cases than  students have a 0\% of isolation, reflecting in \mbox{32,817} fewer deaths.
Although 0\% and 100\% isolation levels may be infeasible, there is no way to predict which isolation level would be achieved.
If we compare an 80\% isolation to a 20\% isolation, we reduce the number of deaths by up to \mbox{16,844}.
While comparing a range from 60\% isolation to a 40\% isolation, we still reduce the number of deaths significantly, by up to \mbox{4,748}.
Even with a 0\% isolation, although the amount of deaths is higher, the maximum number of required ICUs was \mbox{5,212} after schools opened, only a little higher than the capacity of SPMA.
As expected, results show that student behavior is a key factor for the success of this strategy.
The major problem is that there is no way to predict how students are going to behave.

The last strategy we evaluate, shown in Figure~\ref{fig:strategy:vaccine}, is to return to classes and open schools only on day~341, which corresponds to February 1st, 2021.
We consider that all students return to school but follow sanitary measures (masks and hygiene).
For this strategy, we analyze two different scenarios: (1) no vaccines are available; (2) vaccines are available since day~310 (January 1st, 2020).
For the vaccines scenario, the vaccines are administered with a procedure divided into four phases, in the following order: people who are over 50 years old, teachers, students and, finally, the rest of the population.
We consider that \mbox{300,000} people can receive the vaccine per day with 90\% effectiveness rate and the vaccine takes 14 days to generate the individual immune response.

The most important thing to note is that, in the scenario without vaccines, regardless if schools open on day~224 or~341, opening schools with all students at once has a deep impact in the number of cases, deaths and health care system.
On the other hand, with the vaccines, we can observe that schools can be reopened without any major concerns about ICU beds availability.
In this case, up to \mbox{37,753} lives could be saved compared to the previous controlled reopen strategy.
Besides the overall immunity of the population, by prioritizing teachers and students, we are able to cut the transmission chain of COVID-19 inside the schools.

\section*{Conclusions}

By analyzing all strategies, we conclude that opening schools with all students at once is a strategy that imposes a high risk, such that any government that adopts it, regardless if this or next year, should proceed very carefully.
It can lead to a collapse of the health care system and thereby to the need of further quarantine periods, which could have a catastrophic impact on economics.
It is also important to note that the students, due to their age, are the least affected by the opening of schools.
However, they can act as infection vectors, causing massive spreads to more sensitive people, such as their family and teachers.
Adopting a controlled reopening, with several stages, carefully increasing the number of students, it is notorious that it is able to reduce the speed of the spread of the virus, with a maximum ICU usage peak between 48.8\% and 97.8\% depending on how much people that attend the school follow the sanitary measures (considering a 50\% isolation), such that the health care system is able to handle the demand of new cases.
Finally, by opening schools only when vaccination campaigns be available, within the SPMA alone, up \mbox{37,753} lives could be saved compared to a controlled strategy.
Since we are likely within a few months of a vaccine, we consider the last the most appropriate strategy.

\bibliographystyle{science}

\fancyhead{}


\section*{Acknowledgments}
This research was partially supported by a machine allocation on the Kabr\'e supercomputer at the Costa Rica National High Technology Center.
We would like to thank Marco A. Amato for reviewing an earlier version of this manuscript.

\textbf{Financial support}

This work received no direct funding.

\textbf{Conflict of interest}

Conflicts of Interest: None.

\textbf{Data and materials availability}

In case the paper gets accepted for publication, we will fully disclose all data underlying the study in a public repository.

\clearpage

\section*{Figures}

\begin{figure}[!h]
	\centering

	\begin{subfigure}{0.46\linewidth}
	\includegraphics[width=\textwidth]{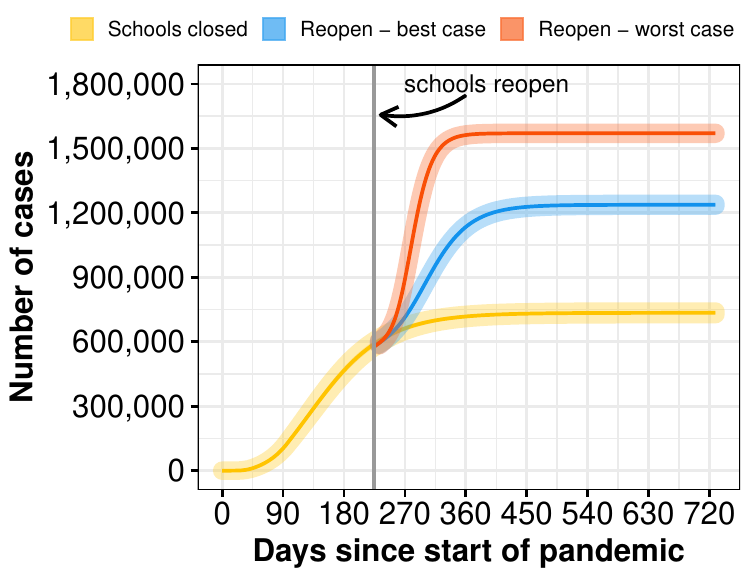}
	\caption{General -- accumulated reported cases.}
	\end{subfigure}
	\hspace{0.05\linewidth}
	\begin{subfigure}{0.46\linewidth}
	\includegraphics[width=\textwidth]{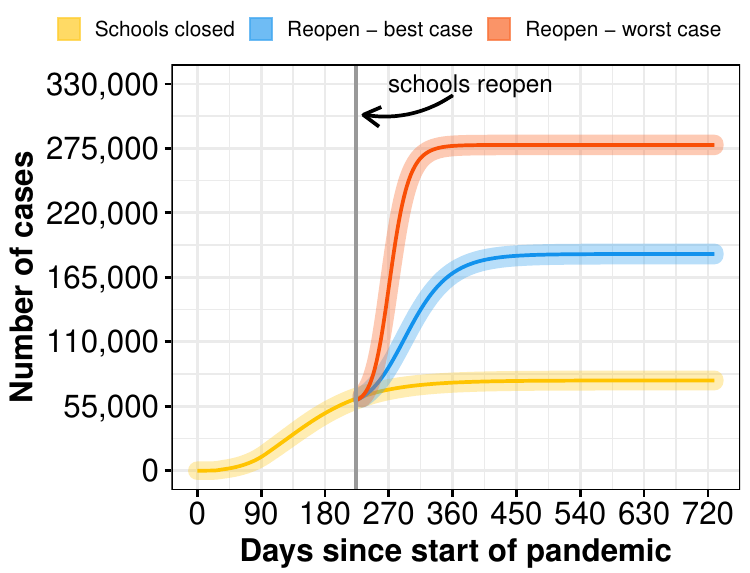}
	\caption{Schools -- accumulated reported cases.}
	\end{subfigure}

	\begin{subfigure}{0.46\linewidth}
	\includegraphics[width=\textwidth]{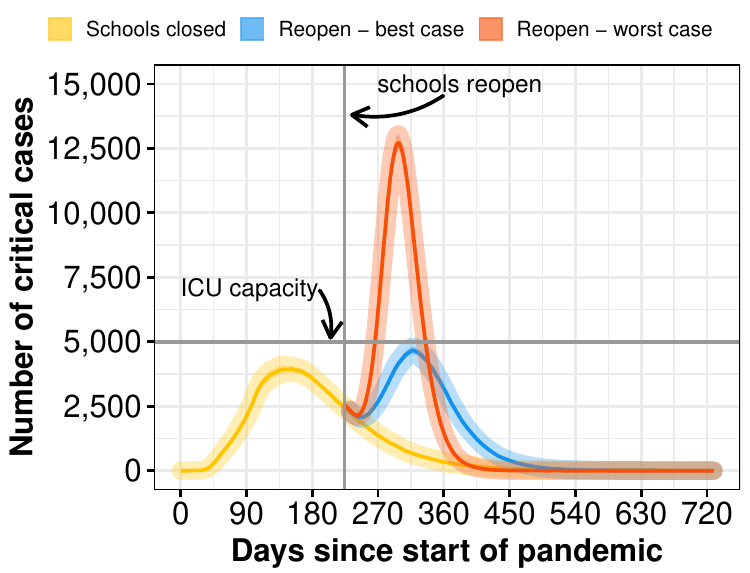}
	\caption{General -- critical cases (ICU).}
	\end{subfigure}
	\hspace{0.05\linewidth}
	\begin{subfigure}{0.46\linewidth}
	\includegraphics[width=\textwidth]{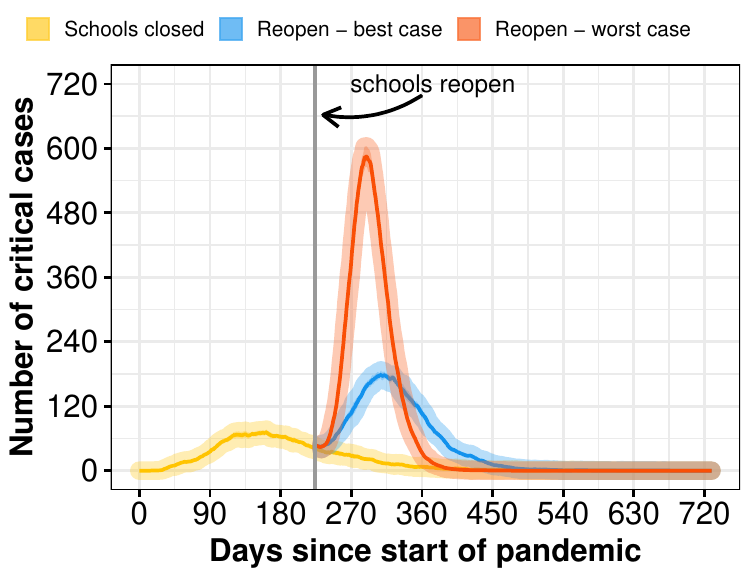}
	\caption{Schools -- critical cases (ICU).}
	\end{subfigure}

	\begin{subfigure}{0.46\linewidth}
	\includegraphics[width=\textwidth]{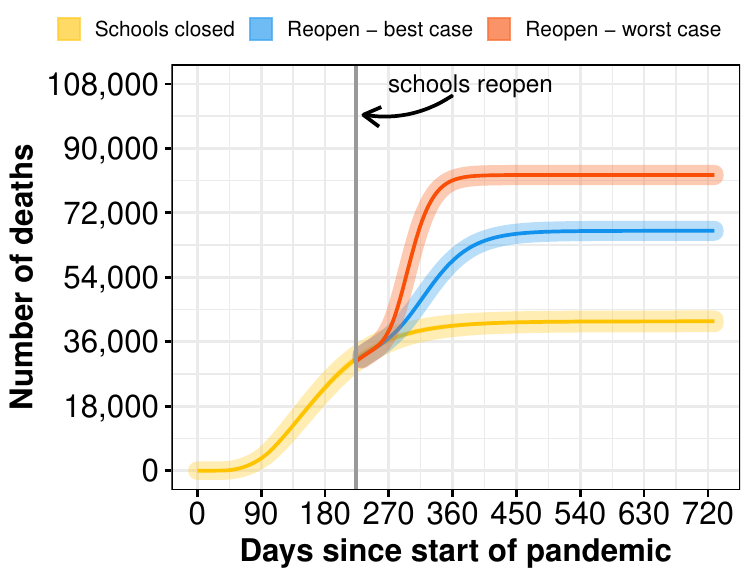}
	\caption{General -- accumulated deaths.}
	\end{subfigure}
	\hspace{0.05\linewidth}
	\begin{subfigure}{0.46\linewidth}
	\includegraphics[width=\textwidth]{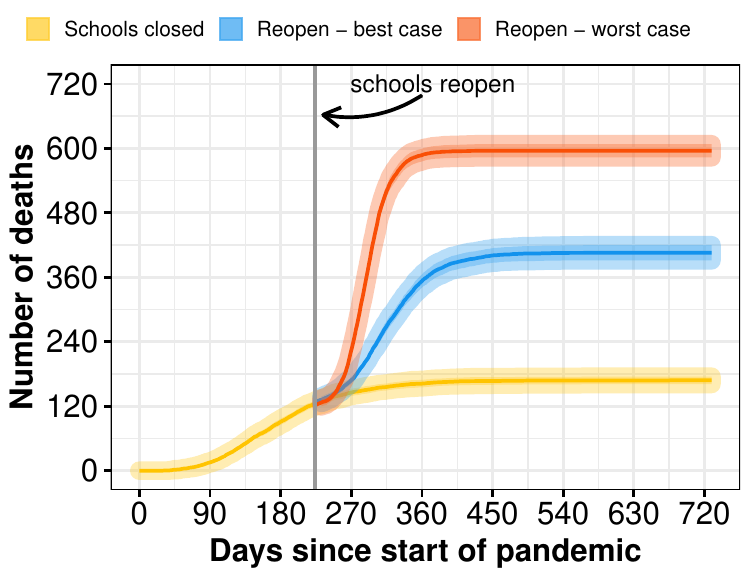}
	\caption{Schools -- accumulated deaths.}
	\end{subfigure}

	\caption{Curves of infected people when schools reopen with all students at once.}
	\label{fig:strategy:all:students}
\end{figure}

\begin{figure}[!h]
	\centering

	\begin{subfigure}{0.44\linewidth}
	\includegraphics[width=\textwidth]{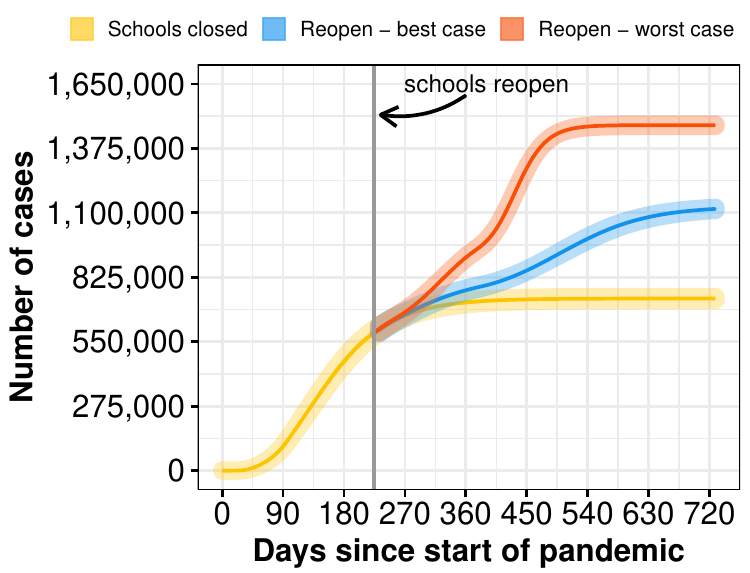}
	\caption{General -- accumulated reported cases.}
	\end{subfigure}
	\hspace{0.05\linewidth}
	\begin{subfigure}{0.44\linewidth}
	\includegraphics[width=\textwidth]{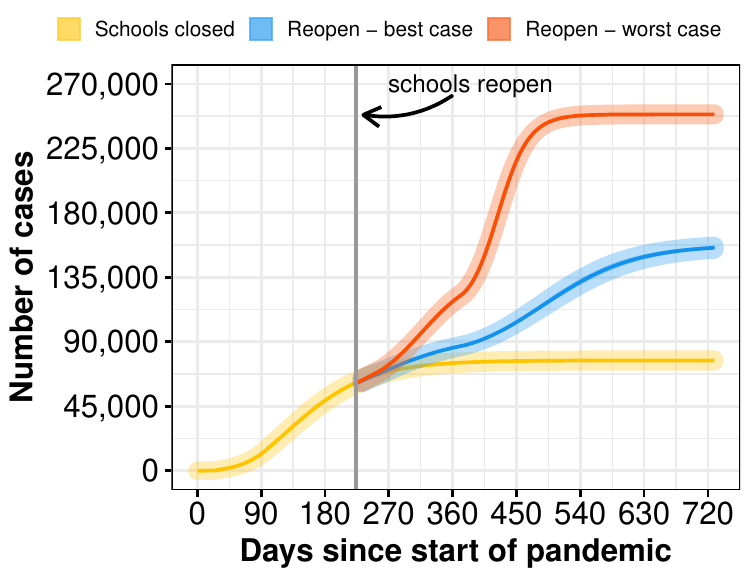}
	\caption{Schools -- accumulated reported cases.}
	\end{subfigure}

	\begin{subfigure}{0.44\linewidth}
	\includegraphics[width=\textwidth]{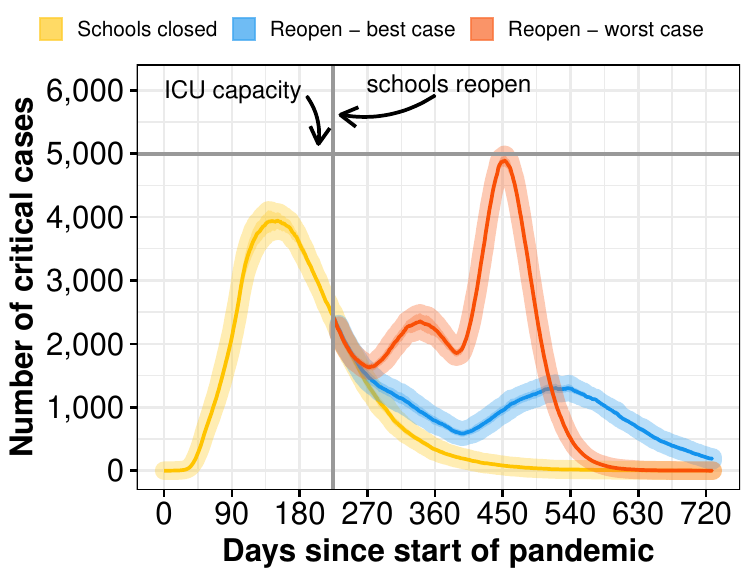}
	\caption{General -- critical cases (ICU).}
	\end{subfigure}
	\hspace{0.05\linewidth}
	\begin{subfigure}{0.44\linewidth}
	\includegraphics[width=\textwidth]{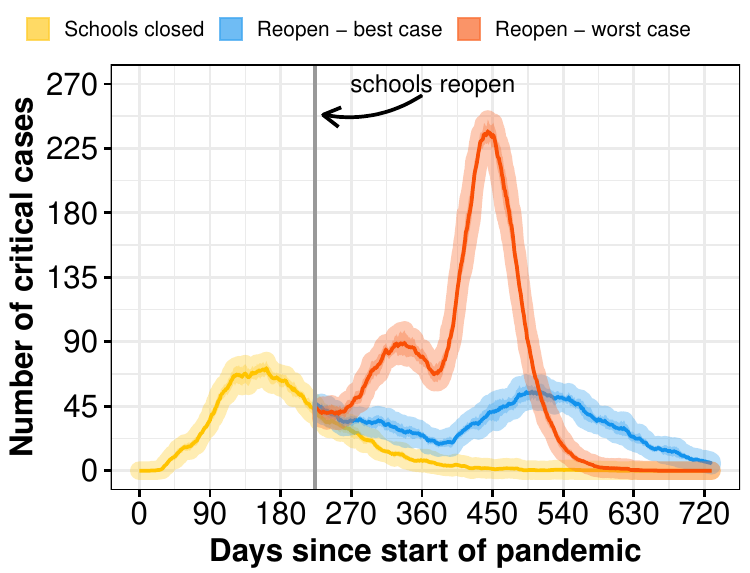}
	\caption{Schools -- critical cases (ICU).}
	\end{subfigure}

	\begin{subfigure}{0.44\linewidth}
	\includegraphics[width=\textwidth]{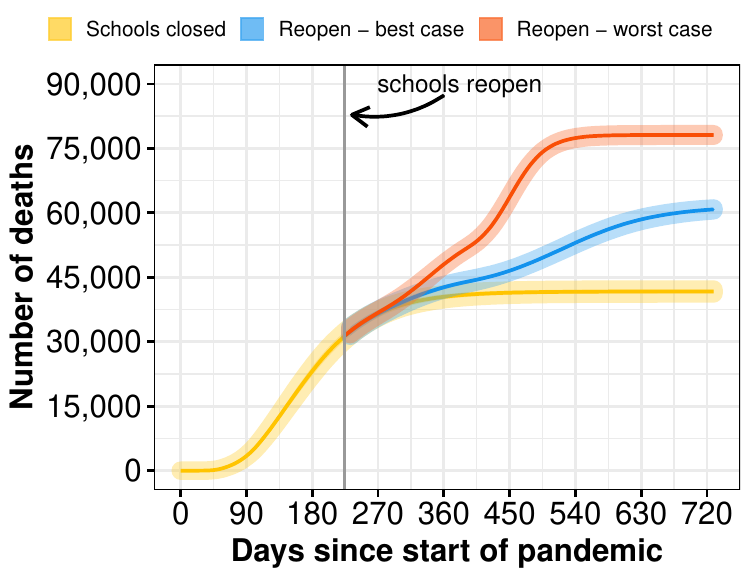}
	\caption{General -- accumulated deaths.}
	\end{subfigure}
	\hspace{0.05\linewidth}
	\begin{subfigure}{0.44\linewidth}
	\includegraphics[width=\textwidth]{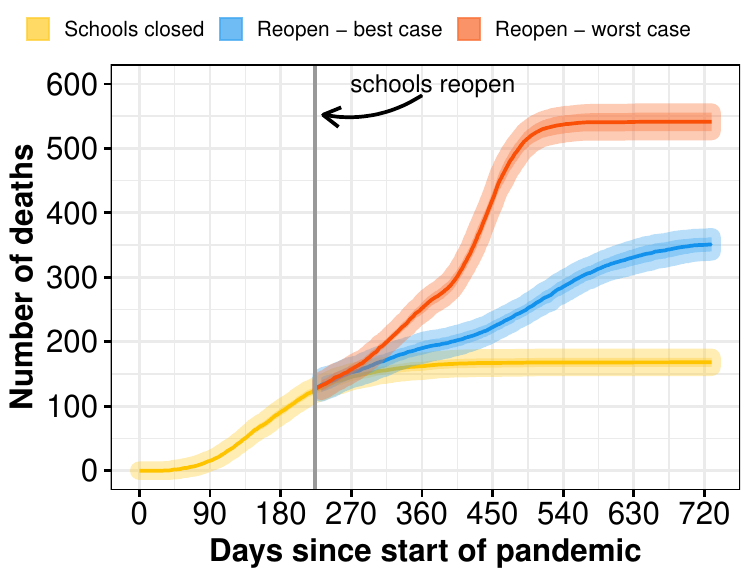}
	\caption{Schools -- accumulated deaths.}
	\end{subfigure}

	\caption{Curves of infected people when schools reopen following São Paulo's strategy.}
	\label{fig:strategy:sp:plan}
\end{figure}

\begin{figure}[!h]
	\centering

	\begin{subfigure}{0.43\linewidth}
	\includegraphics[width=\textwidth]{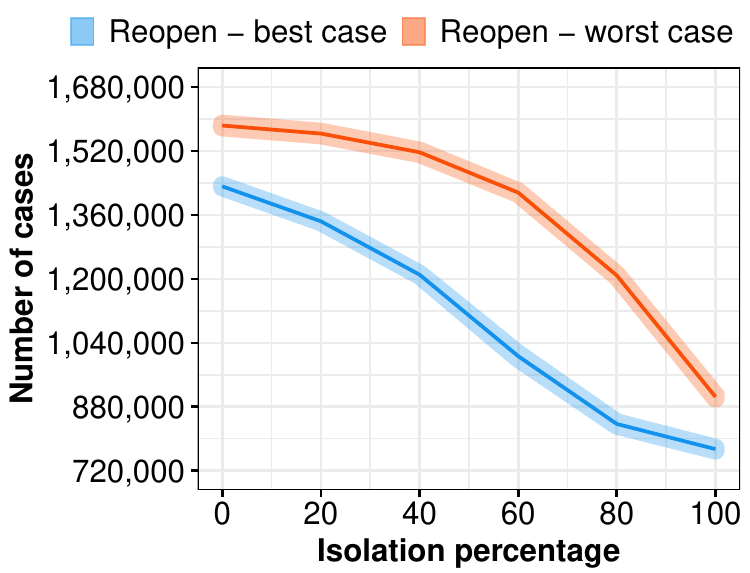}
	\caption{General -- accumulated reported cases.}
	\end{subfigure}
	\hspace{0.05\linewidth}
	\begin{subfigure}{0.43\linewidth}
	\includegraphics[width=\textwidth]{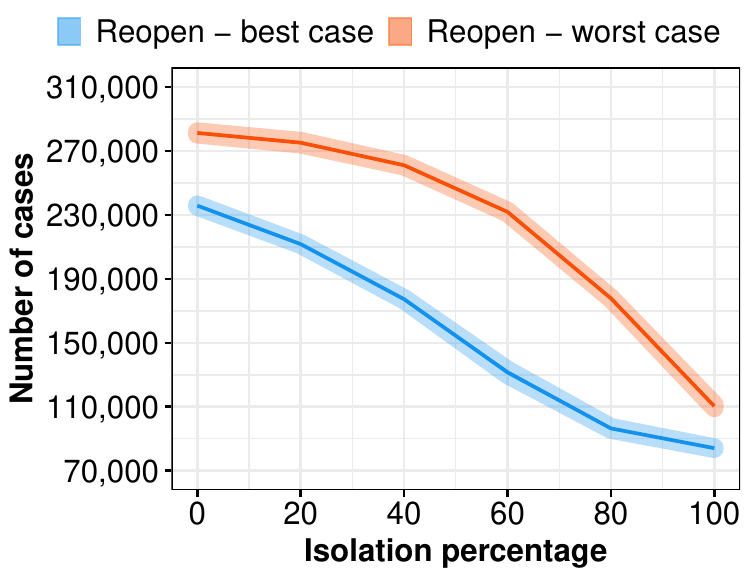}
	\caption{Schools -- accumulated reported cases.}
	\end{subfigure}

	\begin{subfigure}{0.43\linewidth}
	\includegraphics[width=\textwidth]{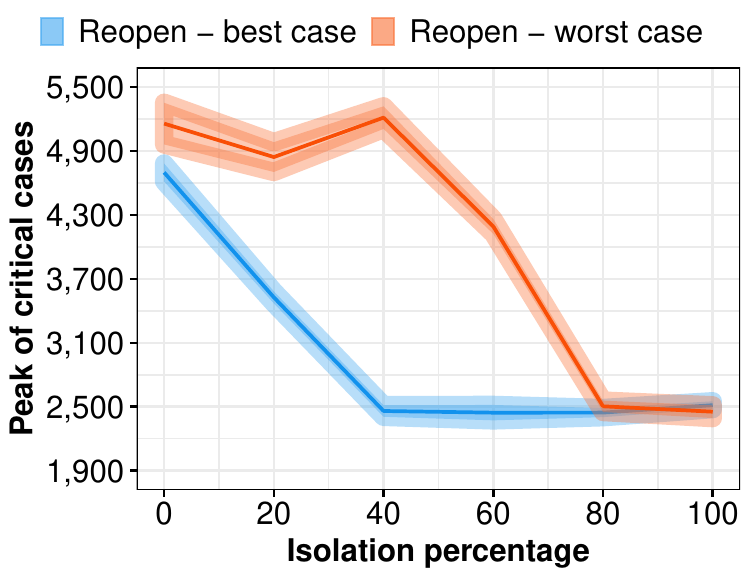}
	\caption{General -- peak of critical cases (ICU) after schools reopen.}
	\end{subfigure}
	\hspace{0.05\linewidth}
	\begin{subfigure}{0.43\linewidth}
	\includegraphics[width=\textwidth]{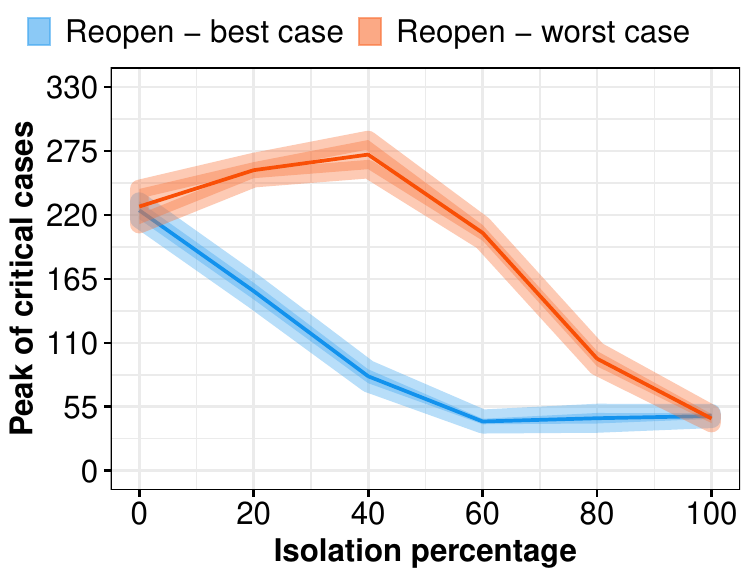}
	\caption{Schools -- peak of critical cases (ICU) after schools reopen.}
	\end{subfigure}

	\begin{subfigure}{0.43\linewidth}
	\includegraphics[width=\textwidth]{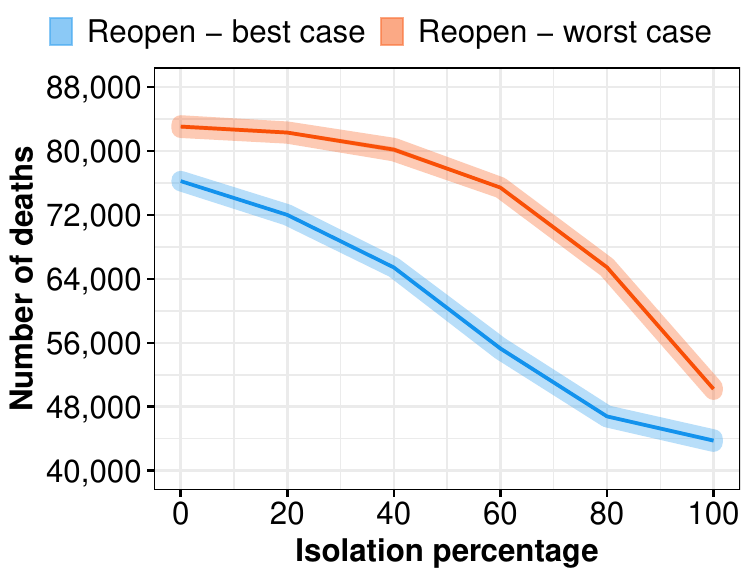}
	\caption{General -- accumulated deaths.}
	\end{subfigure}
	\hspace{0.05\linewidth}
	\begin{subfigure}{0.43\linewidth}
	\includegraphics[width=\textwidth]{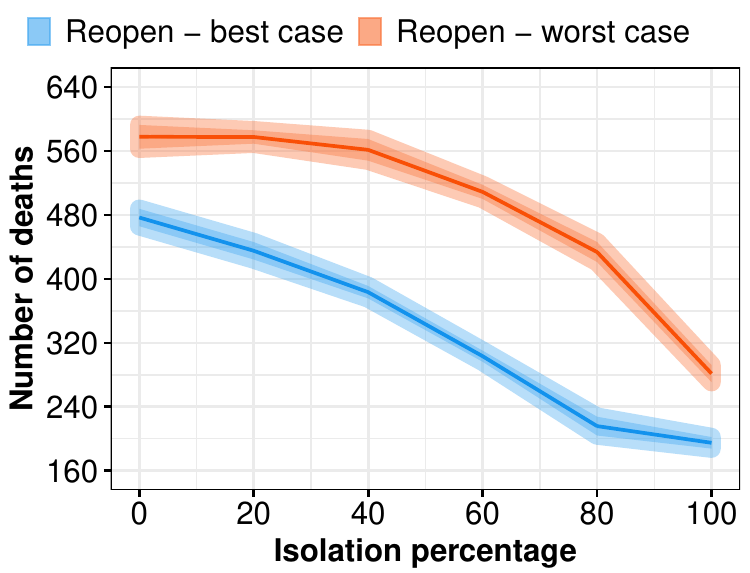}
	\caption{Schools -- accumulated deaths.}
	\end{subfigure}

	\caption{Curves of infected people when schools reopen following São Paulo's strategy, evaluating different isolation levels.}
	\label{fig:strategy:sp:explore}
\end{figure}

\begin{figure}[!h]
	\centering

	\begin{subfigure}{0.46\linewidth}
	\includegraphics[width=\textwidth]{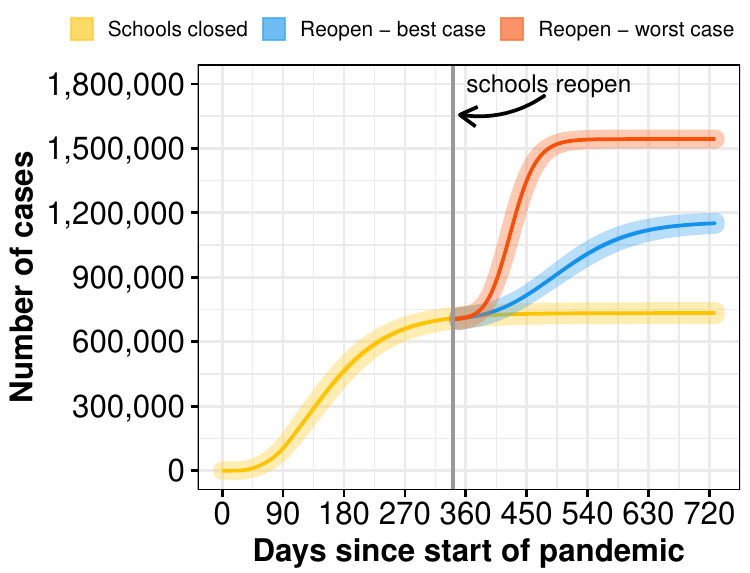}
	\caption{No vaccine -- accumulated reported cases.}
	\end{subfigure}
	\hspace{0.05\linewidth}
	\begin{subfigure}{0.46\linewidth}
	\includegraphics[width=\textwidth]{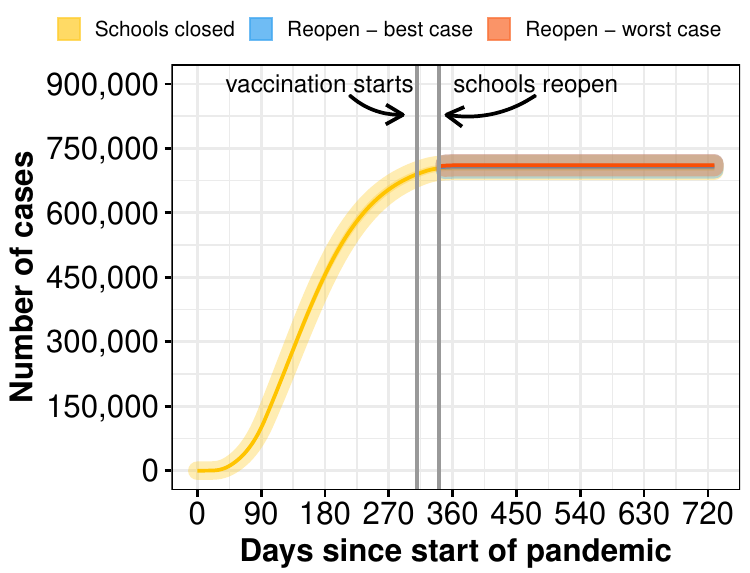}
	\caption{With vaccine -- accumulated reported cases.}
	\end{subfigure}

	\begin{subfigure}{0.46\linewidth}
	\includegraphics[width=\textwidth]{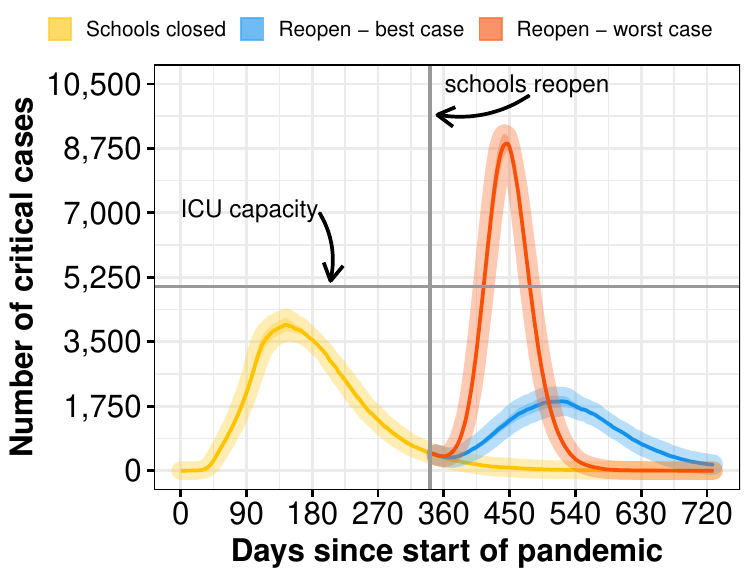}
	\caption{No vaccine -- critical cases (ICU).}
	\end{subfigure}
	\hspace{0.05\linewidth}
	\begin{subfigure}{0.46\linewidth}
	\includegraphics[width=\textwidth]{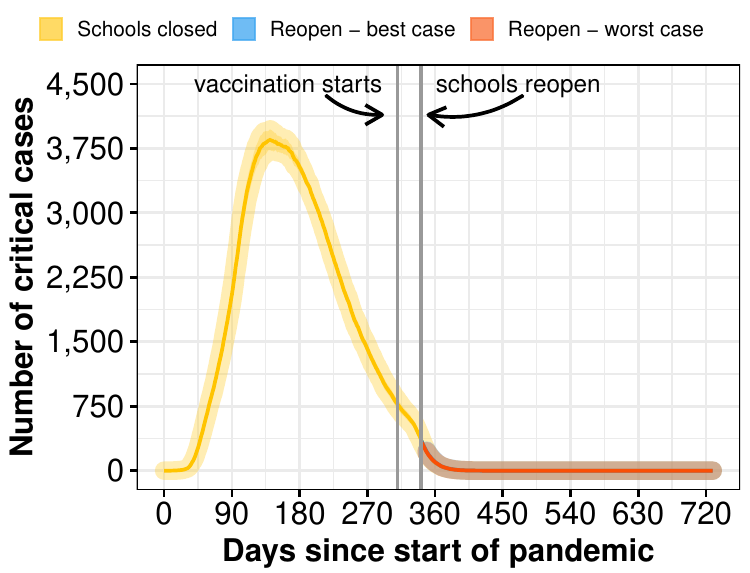}
	\caption{With vaccine -- critical cases (ICU).}
	\end{subfigure}

	\begin{subfigure}{0.46\linewidth}
	\includegraphics[width=\textwidth]{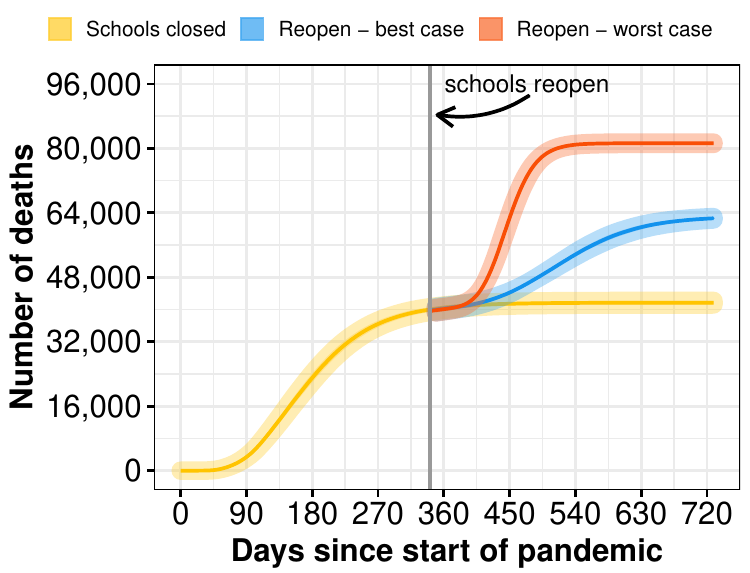}
	\caption{No vaccine -- accumulated deaths.}
	\end{subfigure}
	\hspace{0.05\linewidth}
	\begin{subfigure}{0.46\linewidth}
	\includegraphics[width=\textwidth]{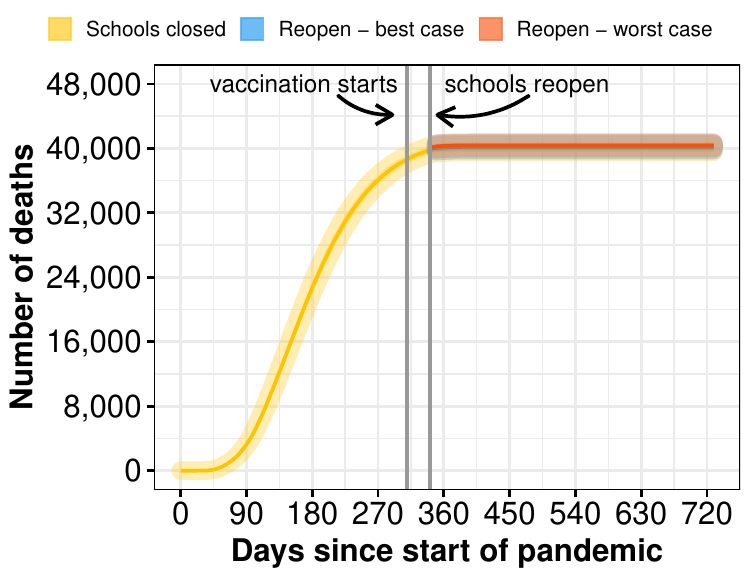}
	\caption{With vaccine -- accumulated deaths.}
	\end{subfigure}

	\caption{Curves of infected people when schools reopen with all students on day~341.}
	\label{fig:strategy:vaccine}
\end{figure}

\clearpage

\vspace{5cm}

{\huge Supplementary Material}

\vspace{2cm}

\textbf{The supplementary Material includes}:
\begin{description}[noitemsep]
 \item Materials and Methods
 \item Figs. 5 to 11
 \item Tables 1 to 4
 \item References
\end{description}

\textbf{Other Supplementary Materials for this manuscript include the following}:
\begin{description}[noitemsep]
 \item Repository address of the simulator source code and input data: Will be disclosed upon paper approval.
\end{description}

\clearpage

\section{Materials and Methods}
\label{sec:methodology}

\subsection{Generating the Relation Network}
\label{sec:methodology:network}

\begin{figure}[H]
\captionsetup[sub]{justification=centering}
	\centering

	\begin{subfigure}{0.46\linewidth}
	\includegraphics[]{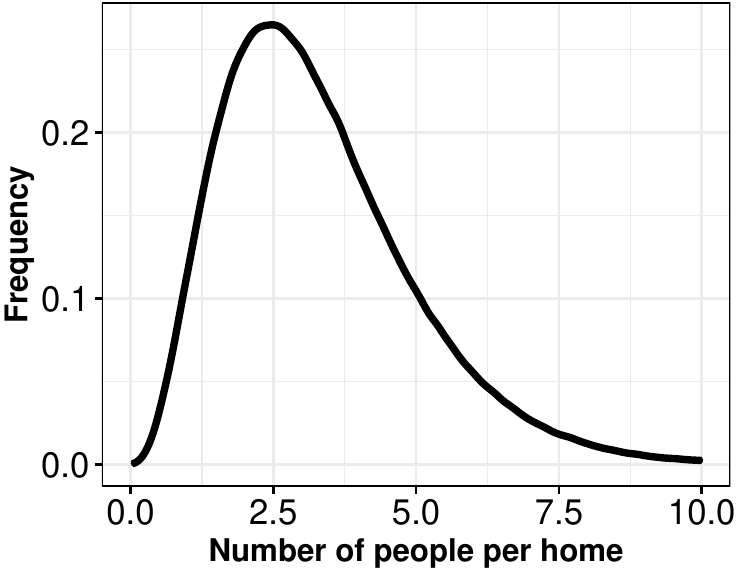}
	\caption{Home.}
	\label{fig:dist:network:home}
	\end{subfigure}
	\hspace{0.05\linewidth}
	\begin{subfigure}{0.46\linewidth}
	\includegraphics[]{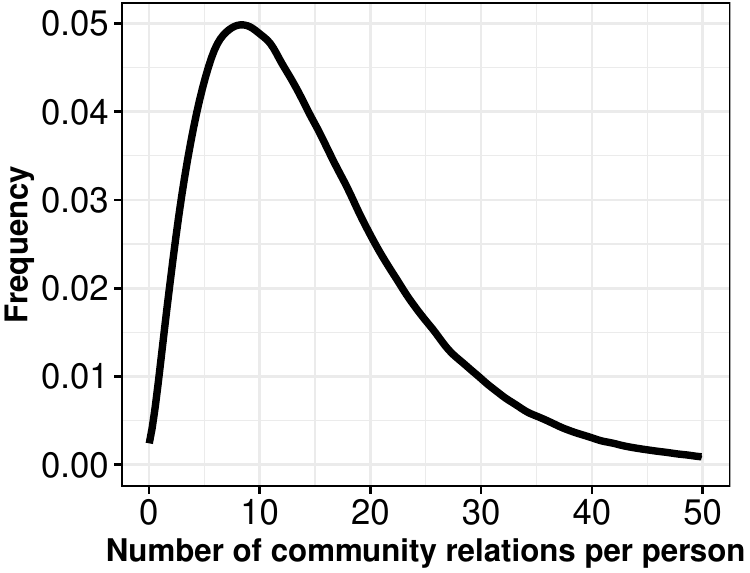}
	\caption{Community.}
	\label{fig:dist:network:community}
	\end{subfigure}

	\begin{subfigure}{0.46\linewidth}
	\includegraphics[]{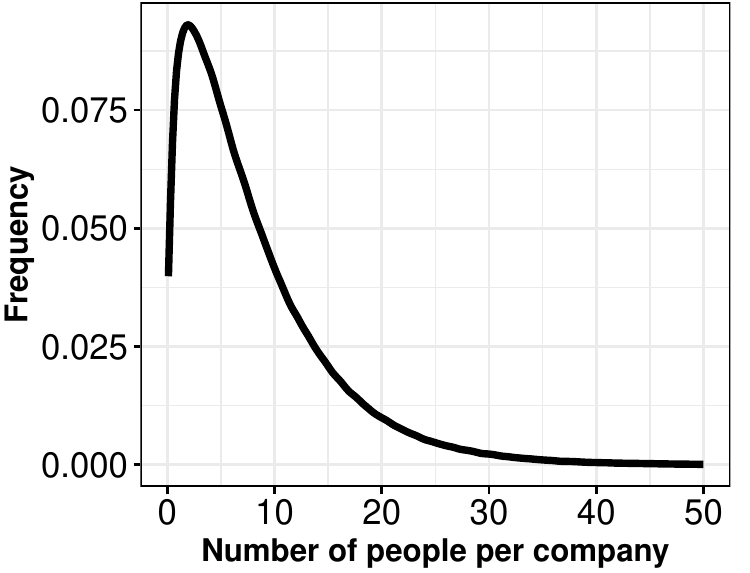}
	\caption{Workplace.}
	\label{fig:dist:network:workplace}
	\end{subfigure}
	\hspace{0.05\linewidth}
	\begin{subfigure}{0.46\linewidth}
	\includegraphics[]{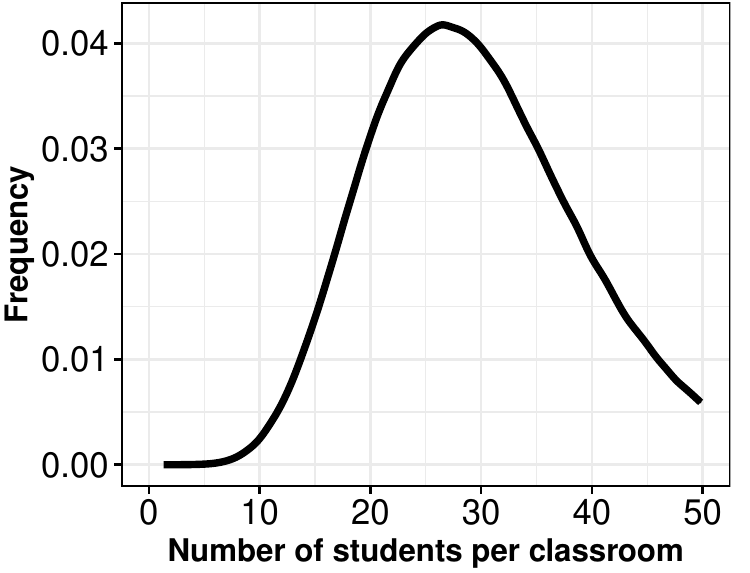}
	\caption{School -- classroom.}
	\label{fig:dist:network:school:classroom}
	\end{subfigure}

	\begin{subfigure}{0.46\linewidth}
	\includegraphics[]{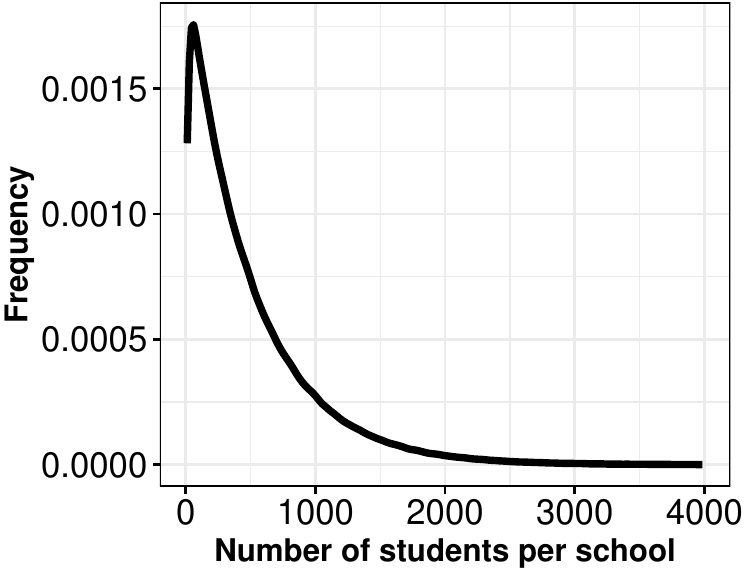}
	\caption{School -- total.}
	\label{fig:dist:network:school:total}
	\end{subfigure}

	\caption{Frequency distributions used to create the network of relations of the population.}
	\label{fig:dist:network}
\end{figure}

The initial network was created in 5 stages.
Note that the initial network considers a normal scenario with no pandemic and interventions.
We explain how we handle the interventions in Section~\ref{sec:methodology:interventions}.
The initial network was created as follows:

\begin{description}[noitemsep]
\item[Home relations] The number of people in the same home follows a distribution described in Table~\ref{tb:dist:network} and Figure~\ref{fig:dist:network:home}, which was derived from the microdata of the last official census from the Brazilian Institute of Geography and Statistics (IBGE) \cite{ibge:censo:2010:microdata}.
Inside the same home, every person is connected to all other people within the same home.

\item[Community relations] The number of relations of each person follows a distribution described in Table~\ref{tb:dist:network} and Figure~\ref{fig:dist:network:community}.

\item[Workplace relations] The number of people per company follows a distribution described in Table~\ref{tb:dist:network} and Figure~\ref{fig:dist:network:workplace}.
We only model as workplace relations jobs that involve agglomeration of people, such as people that work in the commerce, offices, industry and public sectors.
Other jobs, such as people that are autonomous workers or people that work with agriculture, are indirectly modeled in the community relations.
The data regarding the number of employed people and per job type was extracted from IBGE~\cite{ibge:pnad:jan2020, ibge:pnad:jan2020:notice, ibge:censo:2010:microdata}.
In summary, 64\% of people with 18 years old or above work.
Among people that work, 45\% were modeled using workplace relations.
Also, inside the same workplace, we consider that the probability of two people to be connected in the network as 50\%.

\item[School relations] The number of students inside the same class and the number of students within each school follow the distributions described in Table~\ref{tb:dist:network}, Figure~\ref{fig:dist:network:school:classroom} and Figure~\ref{fig:dist:network:school:total}.
For the relations between students from different classrooms, two students have a $0.5\%$ of probability of having a relation in the network.
For the relations between students within the same classroom, two students have a $50\%$ of probability of having relation in the network.
We add one teacher for each classroom, in which the teacher has a relation to all the students of the same class in the network.
According to the census~\cite{ibge:censo:2010:microdata}, $77.8\%$ of people with 18 years old or below frequent the school.

\item[Inter-city relations] The number of people that routinely travel between the different cities relations was gathered from the census~\cite{ibge:censo:2010:microdata}, which indicates a $17.8\%$ of mobility between cities in the metropolitan region of São Paulo.
However, the data does not show the cities, only if the person goes to a different city.
To overcome this issue, we assume that $90\%$ of the traffic from satellite cities go to the capital, and the rest is divided between the other cities in proportion to their population size.
Regarding the traffic from the capital to satellite cities, the traffic is divided proportionally to their population size.
\end{description}

\subsection{Handling Interventions}
\label{sec:methodology:interventions}

To handle interventions, we can modify the infection rates, as well as add and remove edges from the graph.
Different types of intervention require different modifications.
We handled the interventions evaluated in the paper as follows:

\begin{description}[noitemsep]

\item[Quarantine]
To implement the quarantine, we first reduce the infection rates inside schools to zero. Afterwards, we remove the workplace relations corresponding to 50\% of the workplaces.
The 50\% of the relations that are left correspond to essential services that can not be interrupted.
Finally, we reduce the infection rate of community, workplace and inter-city relations to match the reduced reproduction rate, which also simulates other sanitary measures such as wearing masks.
Home relations are not affected by quarantines.

\item[Re-opening of economic activities]
During the quarantine, we zeroed the infection rate of half of the workplaces. To re-open the economic activities, we just increase the infection rate of such edges.

\item[Re-opening schools with all students]
We only increase the infection rate of school relations, which was zeroed by the quarantine.

\item[Re-opening schools with the São Paulo government strategy]
We first remove all school relations from the graph, backing up the data regarding which students and teachers belonged to the same classroom.
We create 3 more relation types to represent each group of students.
Afterwards, we divide each classroom in 3 and re-create the edges using the newly created relation types.
To select which students go to school in a day, we just need to adjust the infection rates of the new relations, zeroing the relations corresponding to the students that will not go to school that day.

\item[Vaccine]
To simulate the vaccines, we only need to change the state of the person being vaccinated to the immune state, carefully considering that it takes some time to the vaccine to make effect, and that not everyone that gets vaccinated will be immunized.

\end{description}

\subsection{Other Simulations Parameters}
\label{sec:methodology:others}

\begin{figure}[H]
	\captionsetup[sub]{justification=centering}
	\centering

	\begin{subfigure}{0.46\linewidth}
	\includegraphics[]{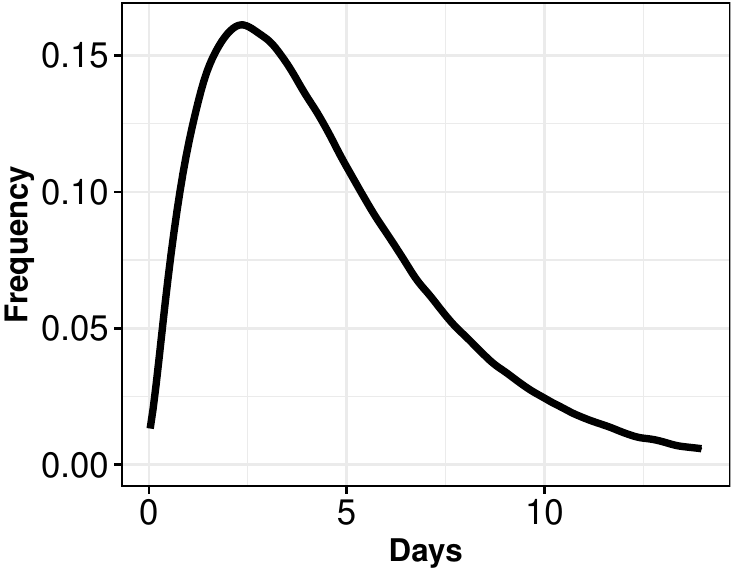}
	\caption{COVID-19 incubation time.}
	\label{fig:params:dist:incubation}
	\end{subfigure}
	\hspace{0.05\linewidth}
	\begin{subfigure}{0.46\linewidth}
	\includegraphics[]{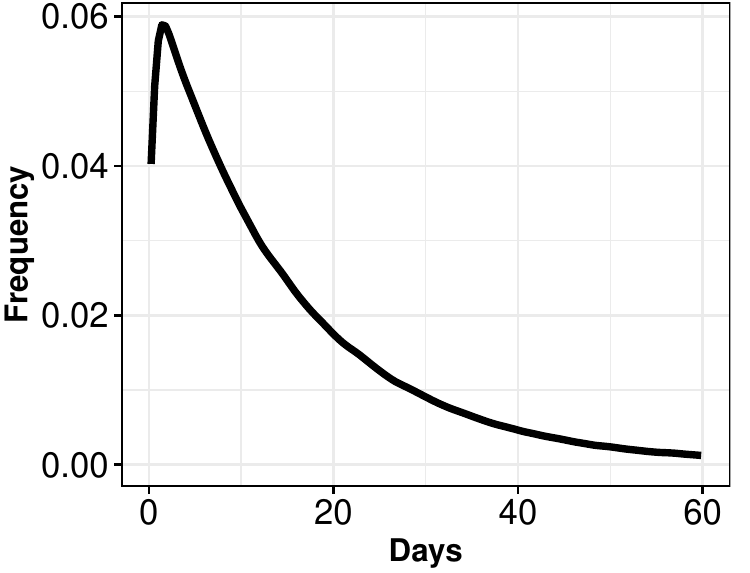}
	\caption{ICU length of stay (critical cases).}
	\label{fig:params:dist:icu:los}
	\end{subfigure}

	\begin{subfigure}{0.46\linewidth}
	\includegraphics[]{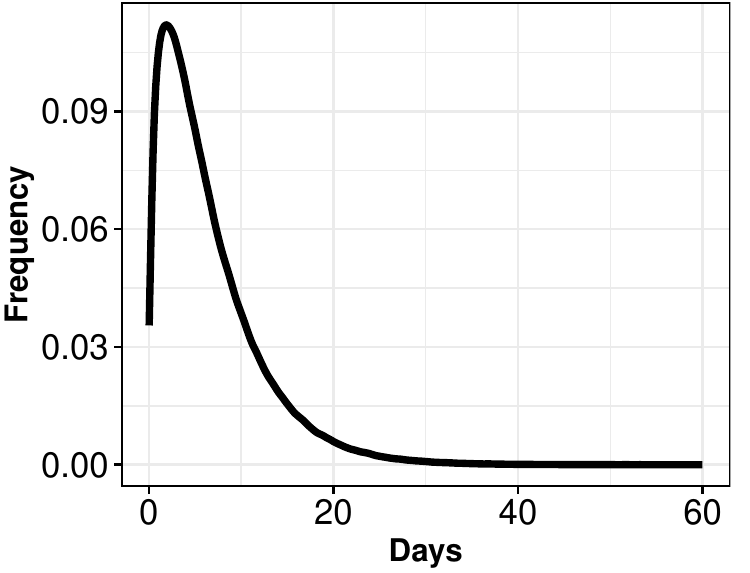}
	\caption{Infirmary length of stay (severe cases).}
	\label{fig:params:dist:infirmary:los}
	\end{subfigure}
	\hspace{0.05\linewidth}
	\begin{subfigure}{0.46\linewidth}
	\includegraphics[]{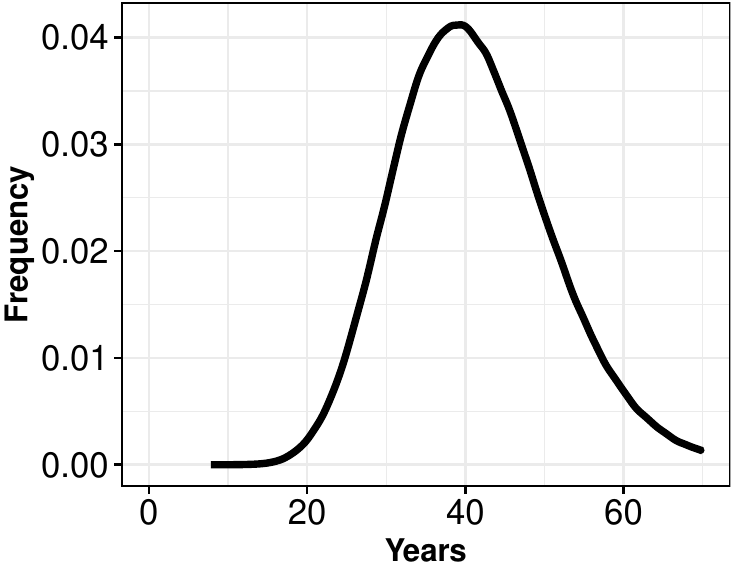}
	\caption{Age of teachers.}
	\label{fig:params:dist:age:teachers}
	\end{subfigure}

	\caption{Frequency distributions of the parameters of Table~\ref{tb:prob:other:params}.}
	\label{fig:params:dist}
\end{figure}

Several simulation parameters can be found in Table~\ref{tb:prob:other:params}~\cite{ibge:pnad:jan2020, ibge:pnad:jan2020:notice, ibge:censo:2010:microdata, sus:datasus:tabnet, Rees2020, PHUA:LANCETRESPIRATORYMEDICINE:2020, sus:opendata}.
The frequency distributions found in the table can be seen in Figure~\ref{fig:params:dist}.

In Table~\ref{tb:prob:symptoms}, we present the probabilities of an infected person to develop symptoms per-age.
The biggest challenge to generate this information is that there is no official data regarding the number of unreported cases per age, as expected by the own definition of unreported.
Therefore, we need to make several transformations with the reported data to take into account the unreported cases in the simulation.
We now describe how we generated these information of Table~\ref{tb:prob:symptoms}.

We first define some functions and variables that return the input data of the algorithm:

\begin{itemize}[noitemsep]
\item Function $people(age)$ returns the number of people of age $age$ in the entire population, which we extracted from the census~\cite{sus:datasus:tabnet}.

\item Functions $reportedMild(age)$, $reportedSevere(age)$ and $reportedCritical(age)$ should return the number of people with mild, severe and critical symptoms per age, respectively.
We extracted this information from the microdata published by the Health Ministry~\cite{sus:opendata}.

\item Variable $globalRatioUnreported$ is the ratio of unreported cases in the entire population. We set it to 0.85 in this paper, since the vast majority of COVID infections are unreported~\cite{science:li:2020}.
\end{itemize}

Next, we define some variable and functions to be used later.
\begin{gather}
totalReportedMild = \sum_{age=0}^{9} reportedMild(age) \\
totalReportedSevere = \sum_{age=0}^{9} reportedSevere(age) \\
totalReportedCritical = \sum_{age=0}^{9} reportedCritical(age)\\
reported(age) = reportedMild(age) + reportedSevere(age) + reportedCritical(age)\\
totalReported = \sum_{age=0}^{9} reported(age)
\end{gather}

To calculate the probability of a person to be unreported, we first calculate the ratio of reported cases:
\begin{gather}
ratioReportedMild(age) = \dfrac{reportedMild(age)}{reported(age)} \\
ratioReportedSevere(age) = \dfrac{reportedSevere(age)}{reported(age)} \\
ratioReportedCritical(age) = \dfrac{reportedCritical(age)}{reported(age)}
\end{gather}

Then, we use Equation~\ref{eq:prob:unreported}, which finds a value $u$ that forces the mean number of unreported people to the desired value.
In the equation, $weight(age)$ returns the proportion of people of age $age$ that should be unreported. Since there is no data for unreported (as expected by the own definition of unreported), we assume that $weight$ has the same ratio of mild cases per age, as it is the softer symptom recorded by official data.

Find $u$, such that:
\begin{gather}
weight(age) = ratioReportedMild(age) \\
u \cdot \dfrac{\sum_{age=0}^{9} weight(age) \cdot people(age)}{\sum_{age=0}^{9} people(age)} = globalRatioUnreported
\label{eq:prob:unreported}
\end{gather}

After finding $u$, we calculate a temporary ratio of unreported cases by age, just used to calculate the symptomatic ratios:
\begin{align}
tmpRatioUnreported(age) &= weight(age) \cdot u
\end{align}

Now that we know the ratio of unreported cases by age, we can calculate the relative proportion (weight) of symptomatic people:
\begin{gather}
weightMild(age) = ratioReportedMild(age) \cdot (1 - tmpRatioUnreported(age)) \\
weightSevere(age) = ratioReportedSevere(age) \cdot (1 - tmpRatioUnreported(age)) \\
weightCritical(age) = ratioCriticalMild(age) \cdot (1 - tmpRatioUnreported(age))
\end{gather}

Now we define our target averages for the ratio of mild, severe and critical cases relative to the entire population (not only relative to reported cases).
\begin{gather}
targetMildRatio = \dfrac{totalReportedMild}{totalReported} \cdot (1 - globalRatioUnreported) \\
targetSevereRatio = \dfrac{totalReportedSevere}{totalReported} \cdot (1 - globalRatioUnreported) \\
targetCriticalRatio = \dfrac{totalReportedCritical}{totalReported} \cdot (1 - globalRatioUnreported)
\end{gather}

Then, we use Equations~\ref{eq:prob:mild}, \ref{eq:prob:severe} and \ref{eq:prob:critical}, which find values $m$, $s$ and $c$ that force the mean ratio of people to the target value.

Find $m$, such that:
\begin{equation}
m \cdot \dfrac{\sum_{age=0}^{9} weightMild(age) \cdot people(age)}{\sum_{age=0}^{9} people(age)} = targetMildRatio
\label{eq:prob:mild}
\end{equation}

Find $s$, such that:
\begin{equation}
s \cdot \dfrac{\sum_{age=0}^{9} weightSevere(age) \cdot people(age)}{\sum_{age=0}^{9} people(age)} = targetSevereRatio
\label{eq:prob:severe}
\end{equation}

Find $c$, such that:
\begin{equation}
c \cdot \dfrac{\sum_{age=0}^{9} weightCritical(age) \cdot people(age)}{\sum_{age=0}^{9} people(age)} = targetCriticalRatio
\label{eq:prob:critical}
\end{equation}

Finally, we calculate the ratios relative to the entire population:
\begin{gather}
ratioMild(age) = weightMild(age) \cdot m \\
ratioSevere(age) = weightSevere(age) \cdot s \\
ratioCritical(age) = weightCritical(age) \cdot c \\
ratioUnreported(age) = 1 - ratioMild(age) - ratioSevere(age) - ratioCritical(age)
\end{gather}

Our model allows us to put different infection rates depending on the patient symptoms.
For patients with severe and critical symptoms, we reduce the infection rate by 50\% to simulate the more controlled environments inside hospitals.
For any symptomatic patients, we zero the infection rates to school and workplace relations.
For patients with severe and critical symptoms, we zero the infection rates to home relations.

Table~\ref{tb:prob:death} shows the death probability of each person depending on the symptoms and age.

\begin{figure}[H]
	\centering
	\includegraphics[]{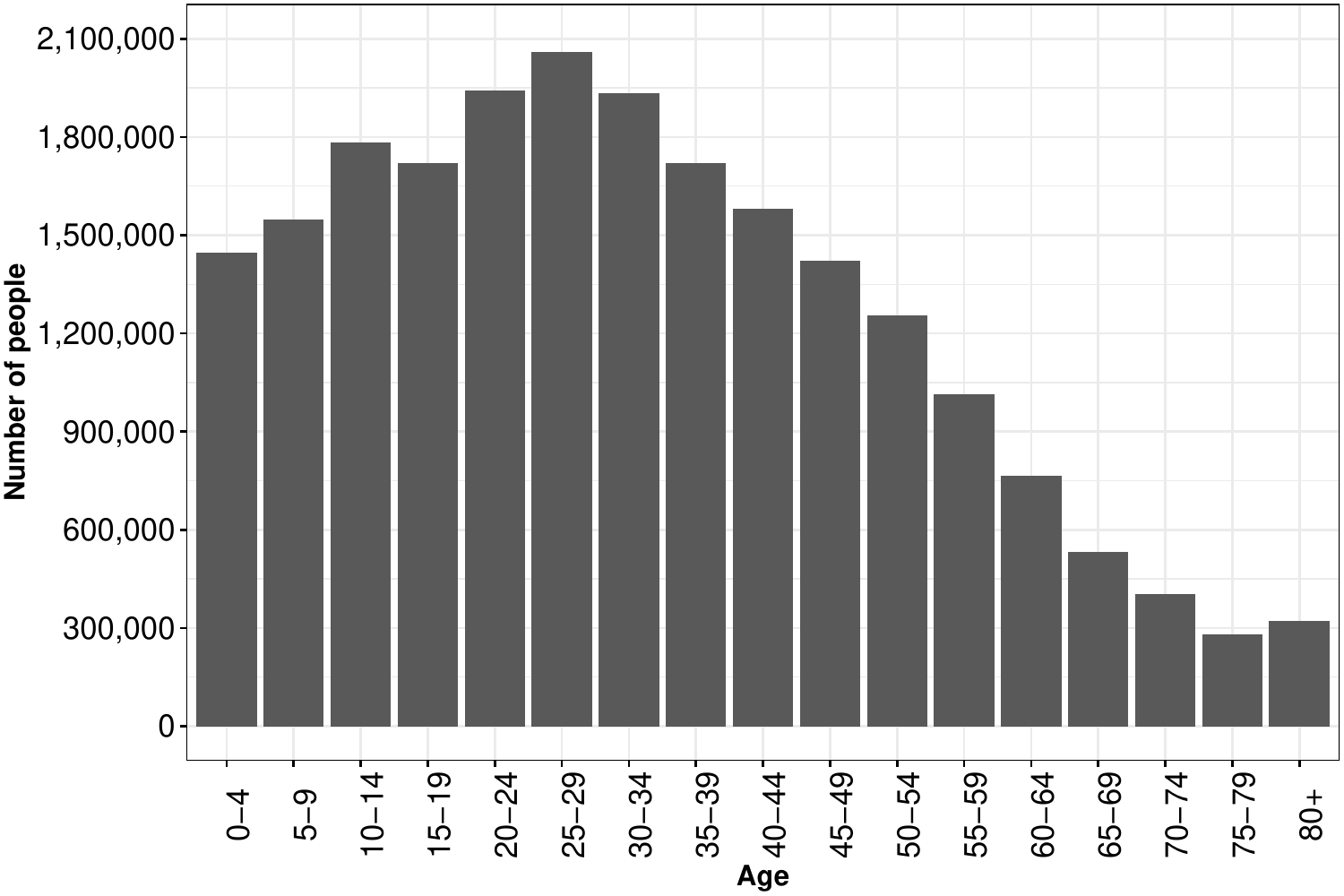}
	\caption{Demographic data of the population of the S\~{a}o Paulo metropolitan area.}
	\label{fig:demographics}
\end{figure}

As we can observe, the age of each person has a deep impact on the symptoms and death rate.
Due to that, we use real demographic data from the census~\cite{sus:datasus:tabnet}.
Since the last census is from 2010, we scale the number of people per age such that the total number of people corresponds to the estimated population size in 2019~\cite{sus:datasus:tabnet}.
The demographic data of the entire São Paulo Metropolitan Area can be seen in Figure~\ref{fig:demographics}.

\subsection{Calibrating the Simulation to Match Real-World Behavior}
\label{sec:methodology:calibration}

\begin{figure}[H]
\captionsetup[sub]{justification=centering}
	\centering

	\begin{subfigure}{\linewidth}
	\includegraphics[]{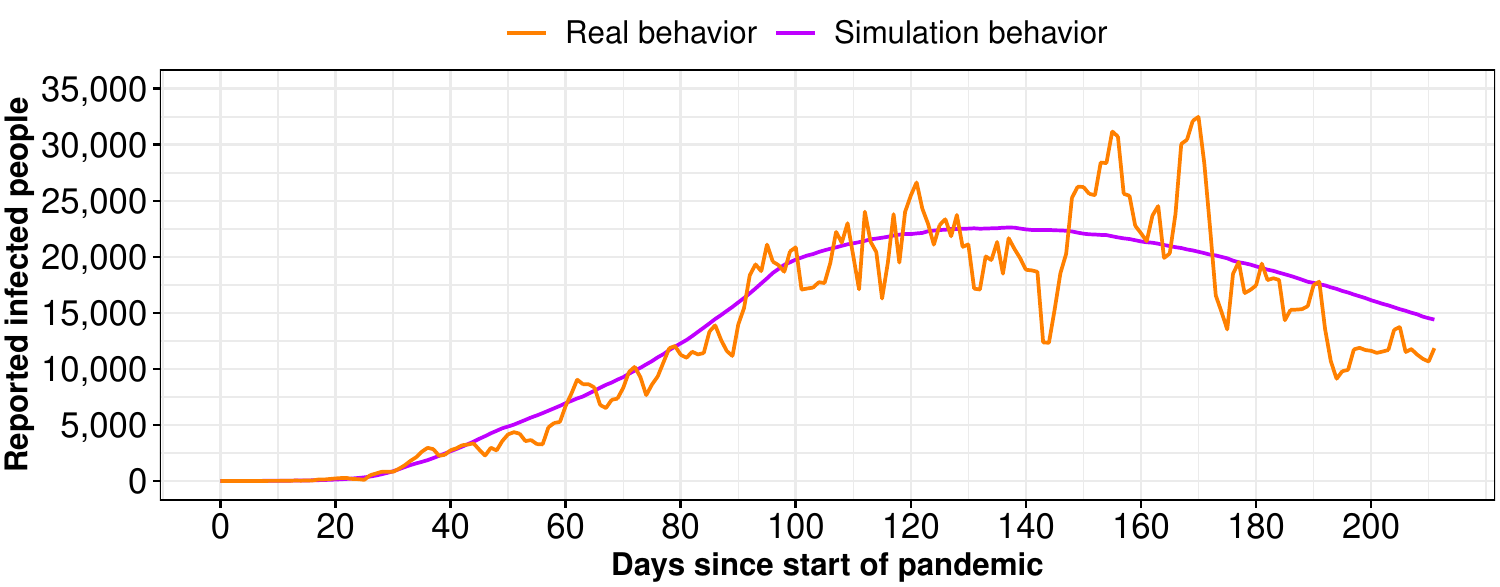}
	\caption{Estimated number of reported infected people.}
	\label{fig:cmpmatch:g}
	\end{subfigure}

	\begin{subfigure}{\linewidth}
	\includegraphics[]{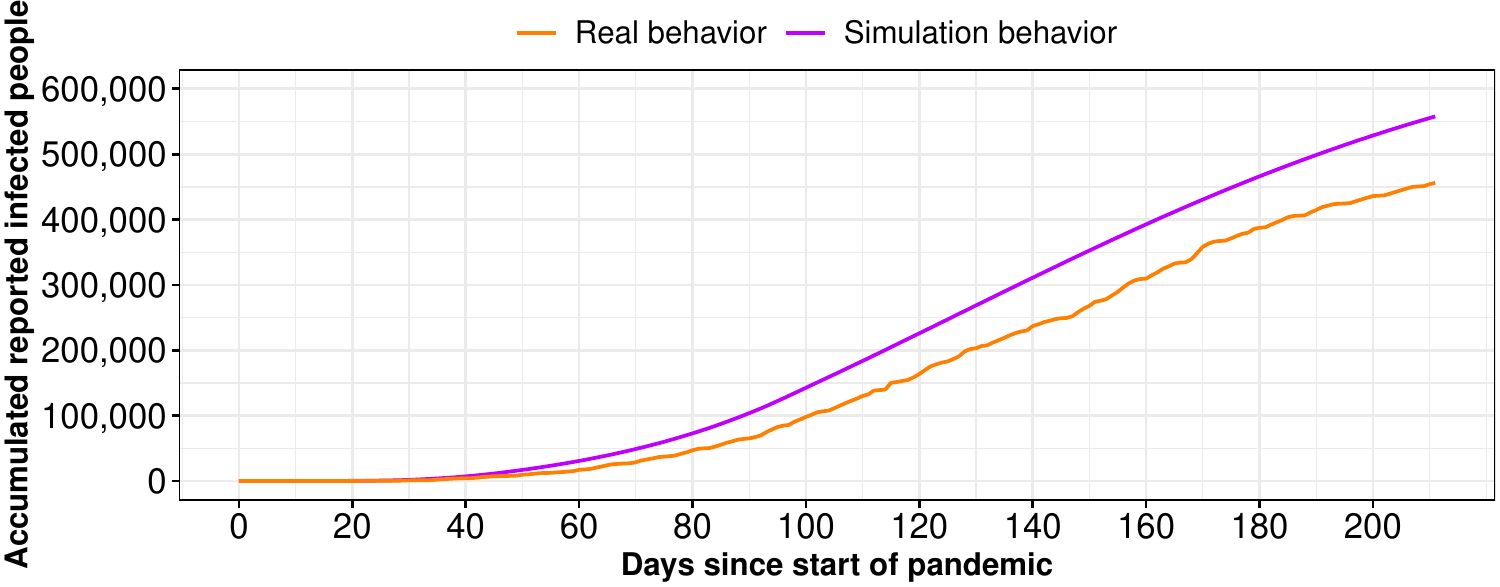}
	\caption{Accumulated number of reported infected people.}
	\label{fig:cmpmatch:ac}
	\end{subfigure}

	\begin{subfigure}{\linewidth}
	\includegraphics[]{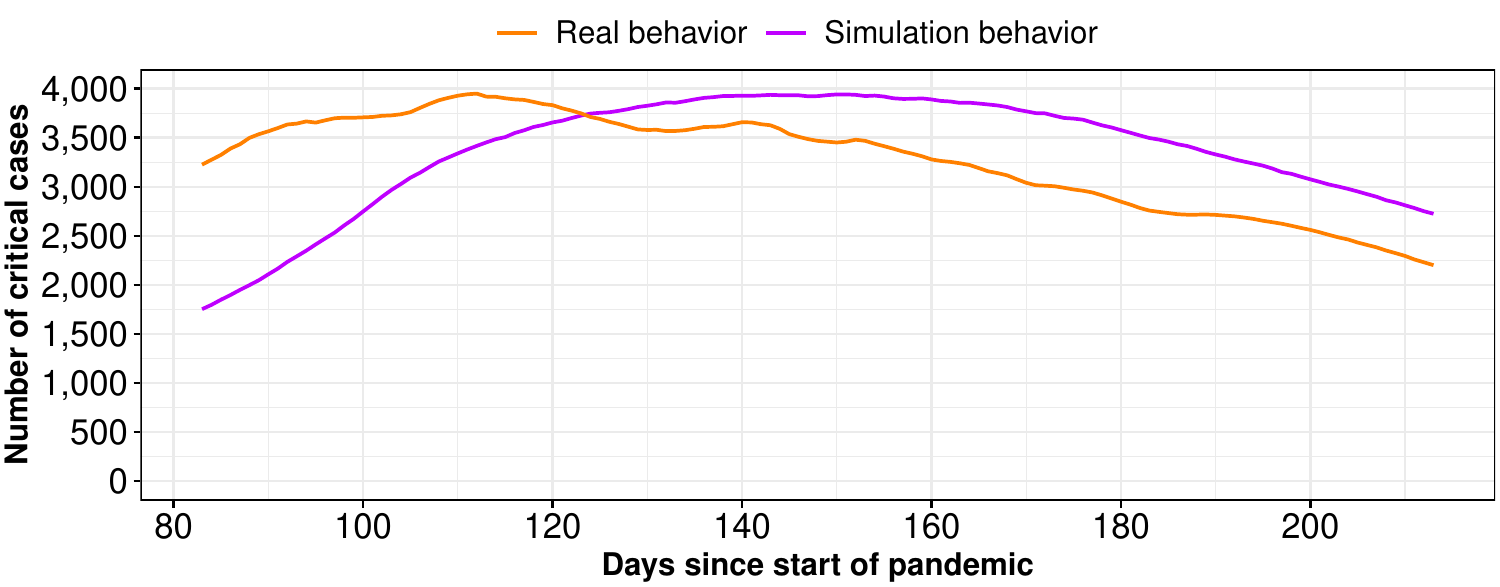}
	\caption{Critical cases -- number of required ICUs.}
	\label{fig:cmpmatch:critical}
	\end{subfigure}

	\caption{Comparison of simulator behavior to the real-world behavior.}
	\label{fig:cmpmatch}
\end{figure}

Figure~\ref{fig:cmpmatch} contains a comparison of the simulator behavior to the real-world data of the São Paulo Metropolitan Area.
The calibration process involves setting the interventions, as explained in Section~\ref{sec:methodology:interventions}, as well as manipulating the infection rates.
In Figure~\ref{fig:cmpmatch:g}, we compare the estimated number of people in the infected state at a day (do not confuse with the number of newly infected people per day).
In Figure~\ref{fig:cmpmatch:ac}, we compare the accumulated number of people in the infected state.
In Figure~\ref{fig:cmpmatch:critical}, we compare the number of critical patients, which represents the ICU occupancy.
It is important to mention that a perfect match between simulator and real-world behavior is impossible, mostly because of the fact that the vast majority of cases are unreported~\cite{science:li:2020}, such that it is impossible to accurately measure real behavior with the official published data.
Nevertheless, we were able to configure the simulation parameters to reflect the real behavior with a reasonable precision, as can be seen in the figures.


\clearpage

\subsection*{Figs. 5 to 11}

\begin{figure}[H]
\captionsetup[sub]{justification=centering}
	\centering

	\begin{subfigure}{0.46\linewidth}
	\includegraphics[]{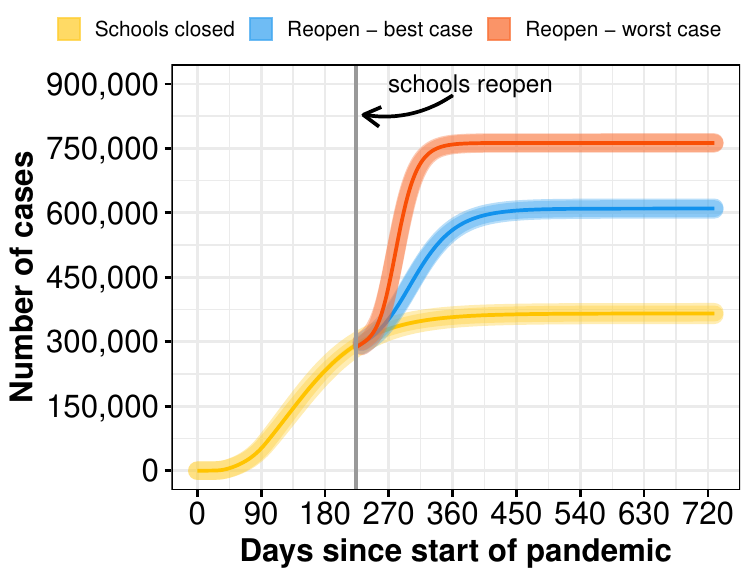}
	\caption{Student's families -- accumulated reported cases.}
	\end{subfigure}
	\hspace{0.05\linewidth}
	\begin{subfigure}{0.46\linewidth}
	\includegraphics[]{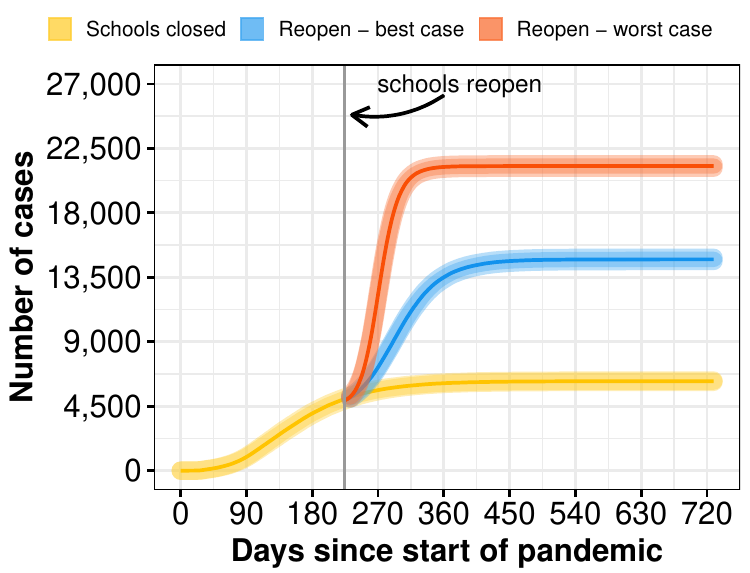}
	\caption{Teachers -- accumulated reported cases.}
	\end{subfigure}

	\begin{subfigure}{0.46\linewidth}
	\includegraphics[]{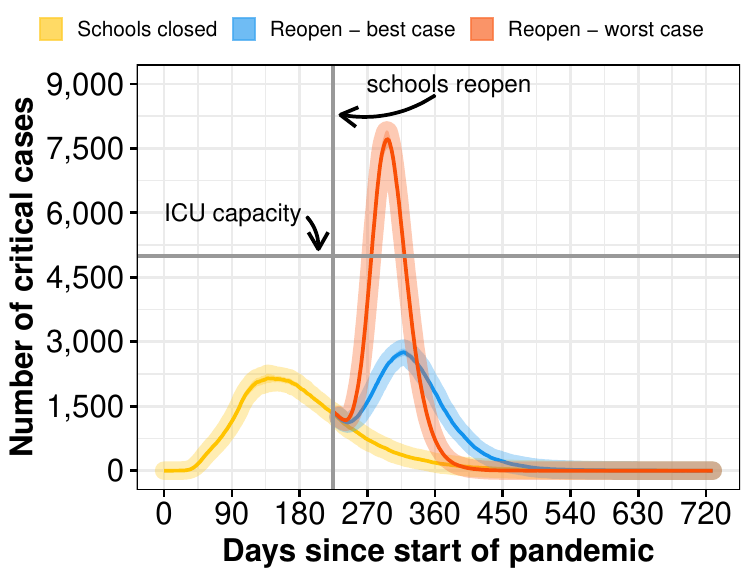}
	\caption{Student's families -- critical cases (ICU).}
	\end{subfigure}
	\hspace{0.05\linewidth}
	\begin{subfigure}{0.46\linewidth}
	\includegraphics[]{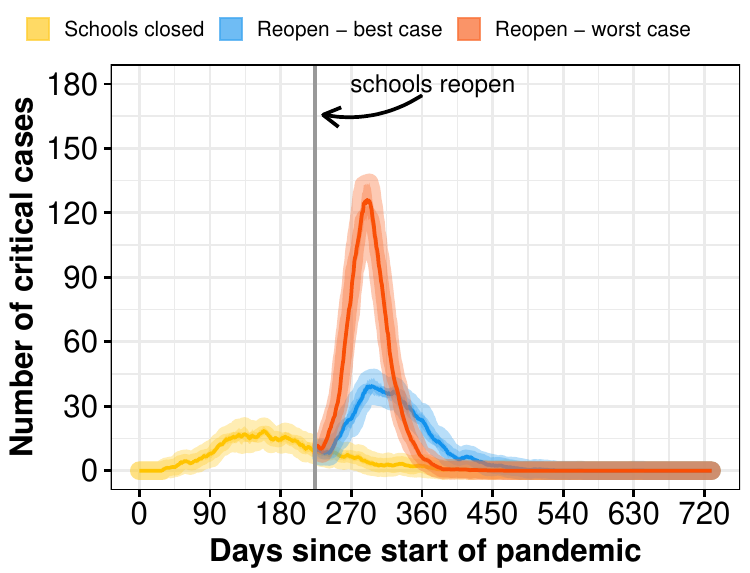}
	\caption{Teachers -- critical cases (ICU).}
	\end{subfigure}

	\begin{subfigure}{0.46\linewidth}
	\includegraphics[]{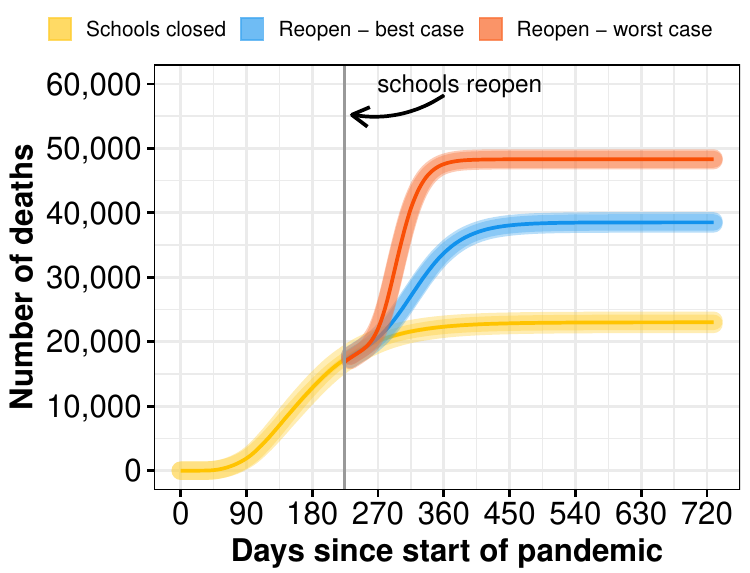}
	\caption{Student's families -- accumulated deaths.}
	\end{subfigure}
	\hspace{0.05\linewidth}
	\begin{subfigure}{0.46\linewidth}
	\includegraphics[]{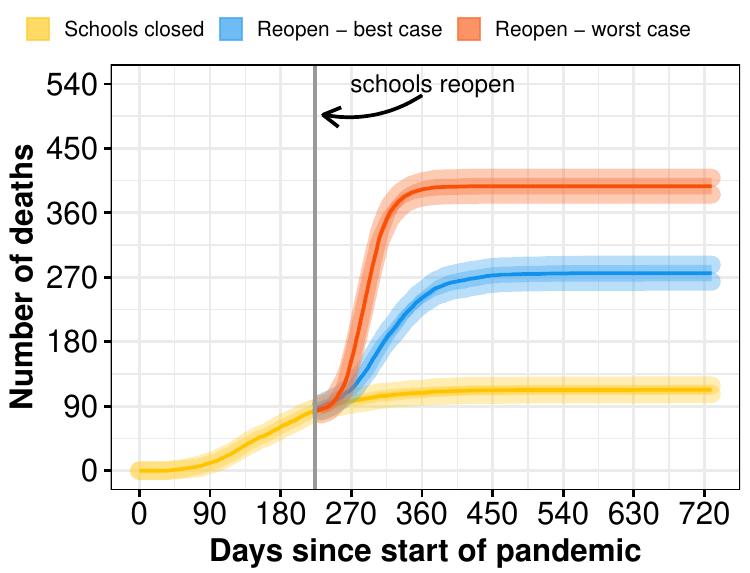}
	\caption{Teachers -- accumulated deaths.}
	\end{subfigure}

	\caption{Curves of infected people when schools reopen with all students at once.}
	\label{fig:strategy:all:students}
\end{figure}

\begin{figure}[H]
\captionsetup[sub]{justification=centering}
	\centering

	\begin{subfigure}{0.46\linewidth}
	\includegraphics[]{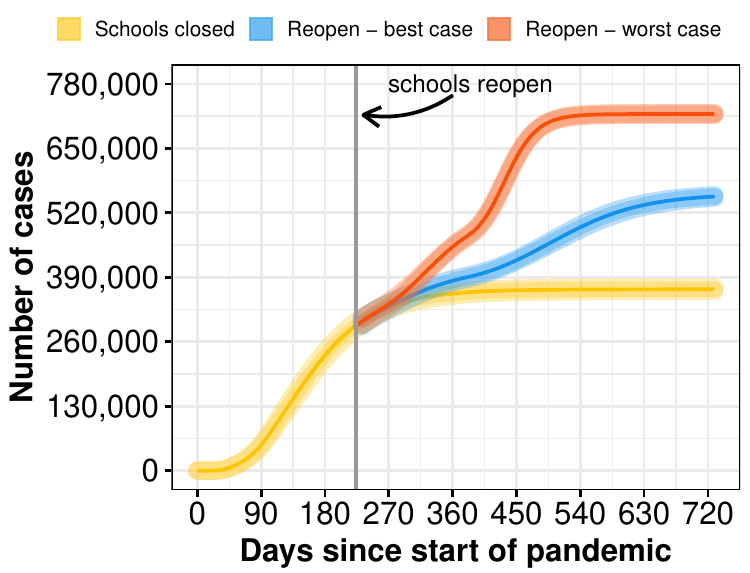}
	\caption{Student's families -- accumulated reported cases.}
	\end{subfigure}
	\hspace{0.05\linewidth}
	\begin{subfigure}{0.46\linewidth}
	\includegraphics[]{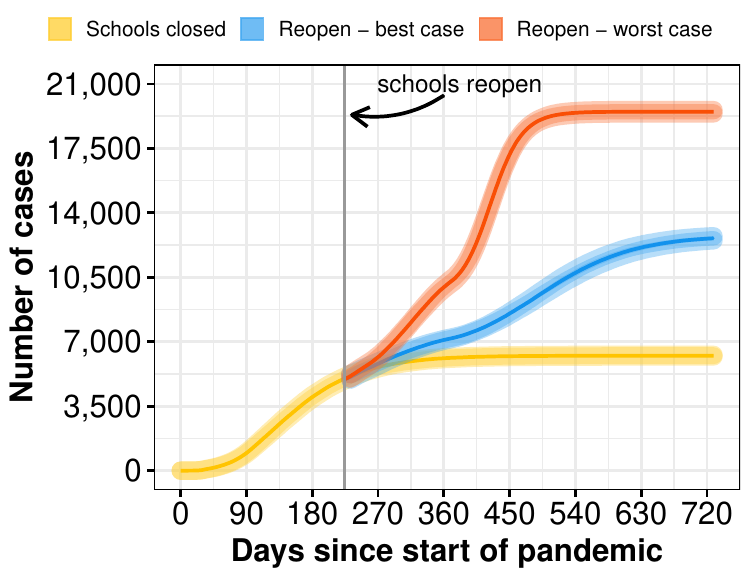}
	\caption{Teachers -- accumulated reported cases.}
	\end{subfigure}

	\begin{subfigure}{0.46\linewidth}
	\includegraphics[]{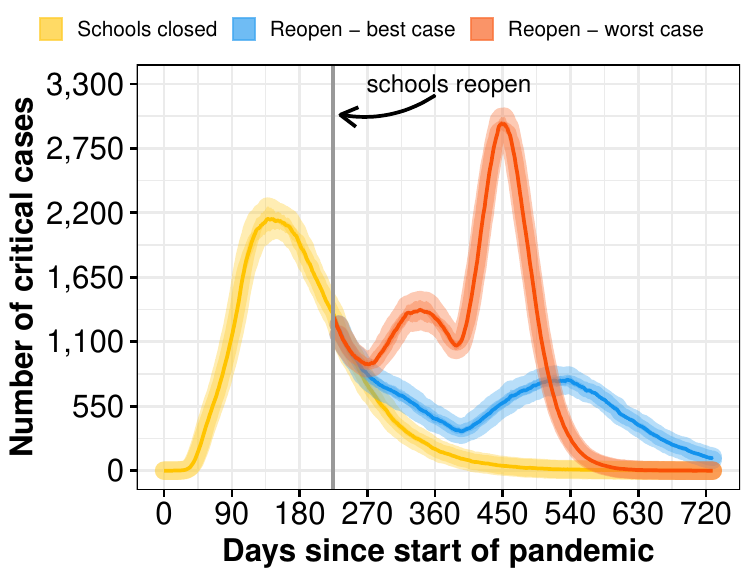}
	\caption{Student's families -- critical cases (ICU).}
	\end{subfigure}
	\hspace{0.05\linewidth}
	\begin{subfigure}{0.46\linewidth}
	\includegraphics[]{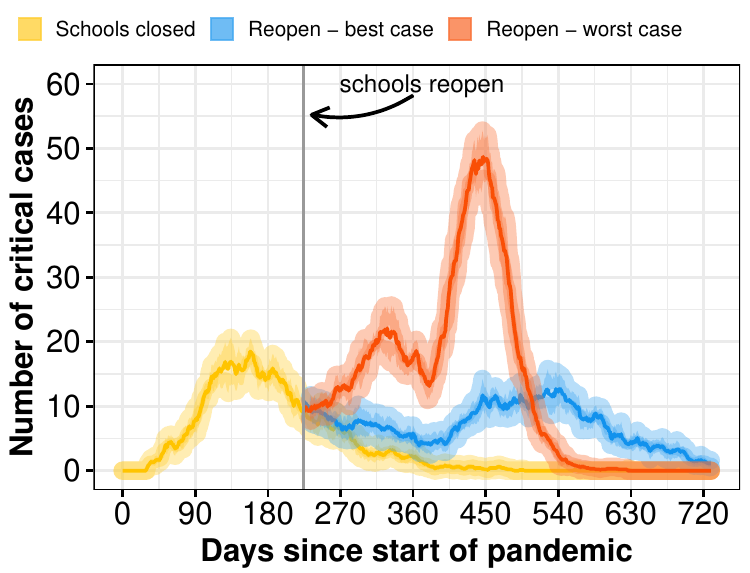}
	\caption{Teachers -- critical cases (ICU).}
	\end{subfigure}

	\begin{subfigure}{0.46\linewidth}
	\includegraphics[]{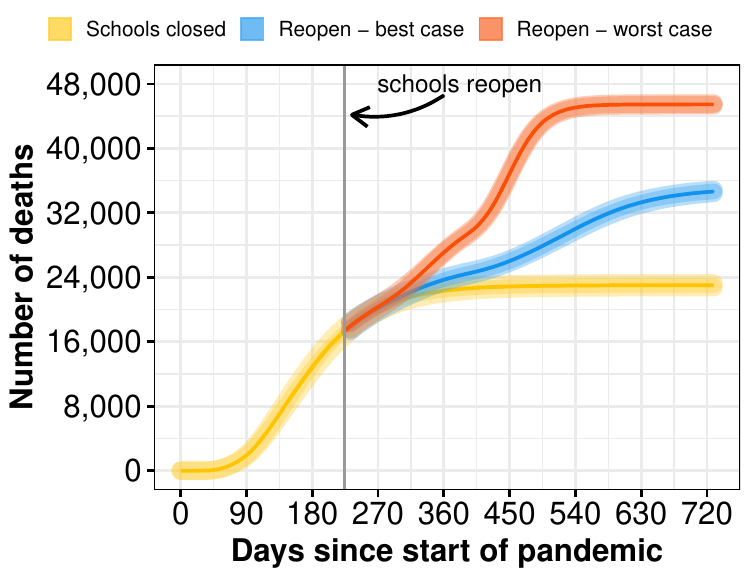}
	\caption{Student's families -- accumulated deaths.}
	\end{subfigure}
	\hspace{0.05\linewidth}
	\begin{subfigure}{0.46\linewidth}
	\includegraphics[]{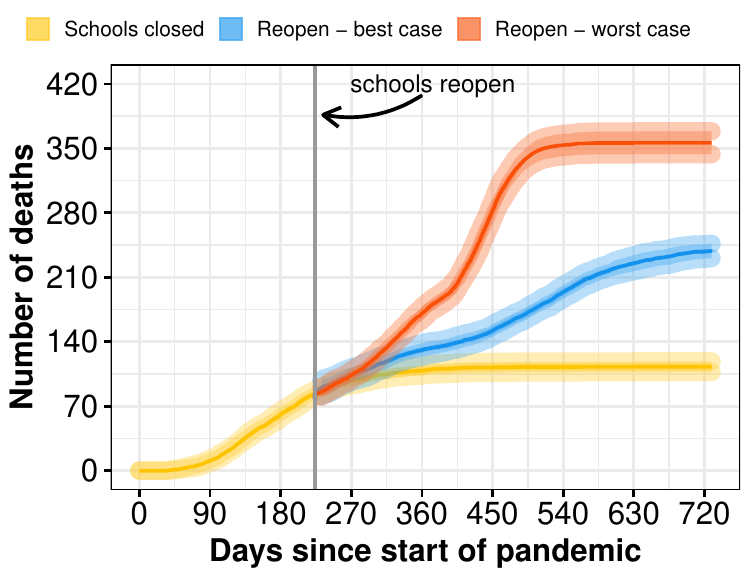}
	\caption{Teachers -- accumulated deaths.}
	\end{subfigure}

	\caption{Curves of infected people when schools reopen following São Paulo's strategy.}
	\label{fig:strategy:sp:plan}
\end{figure}

\begin{figure}[H]
\captionsetup[sub]{justification=centering}
	\centering

	\begin{subfigure}{0.46\linewidth}
	\includegraphics[]{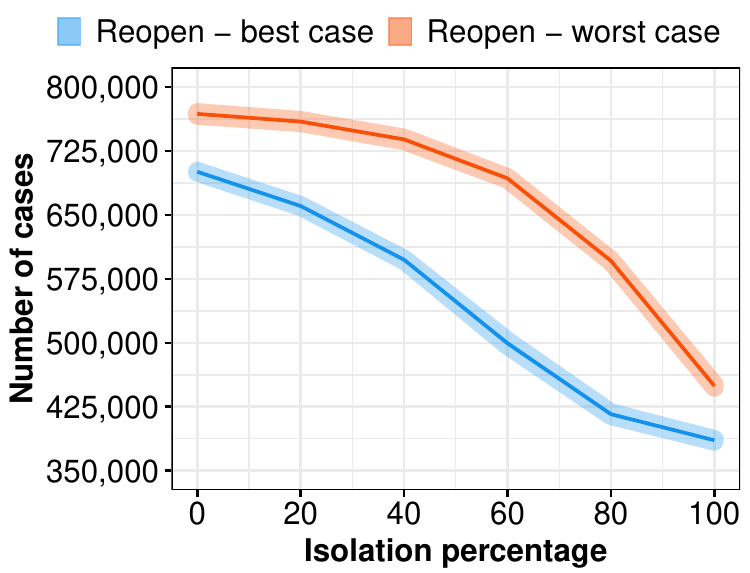}
	\caption{Student's families -- accumulated reported cases.}
	\end{subfigure}
	\hspace{0.05\linewidth}
	\begin{subfigure}{0.46\linewidth}
	\includegraphics[]{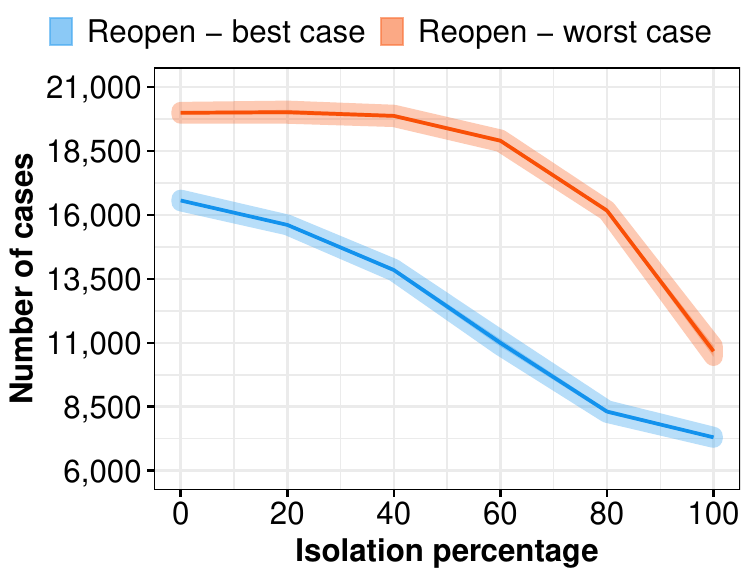}
	\caption{Teachers -- accumulated reported cases.}
	\end{subfigure}

	\begin{subfigure}{0.46\linewidth}
	\includegraphics[]{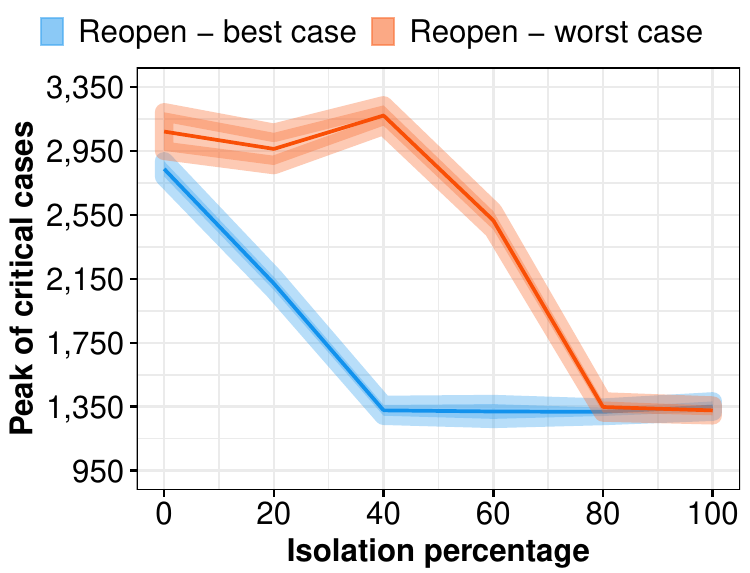}
	\caption{Student's families -- peak of critical cases (ICU) after schools reopening.}
	\end{subfigure}
	\hspace{0.05\linewidth}
	\begin{subfigure}{0.46\linewidth}
	\includegraphics[]{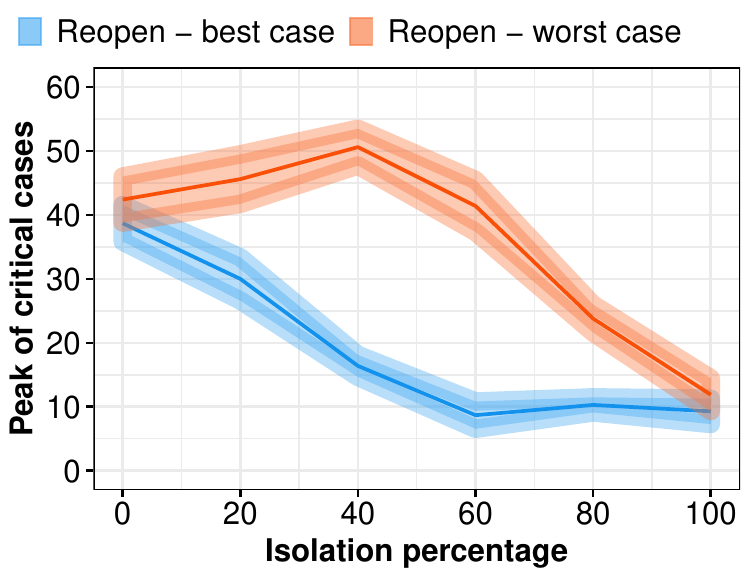}
	\caption{Teachers -- peak of critical cases (ICU) after schools reopening.}
	\end{subfigure}

	\begin{subfigure}{0.46\linewidth}
	\includegraphics[]{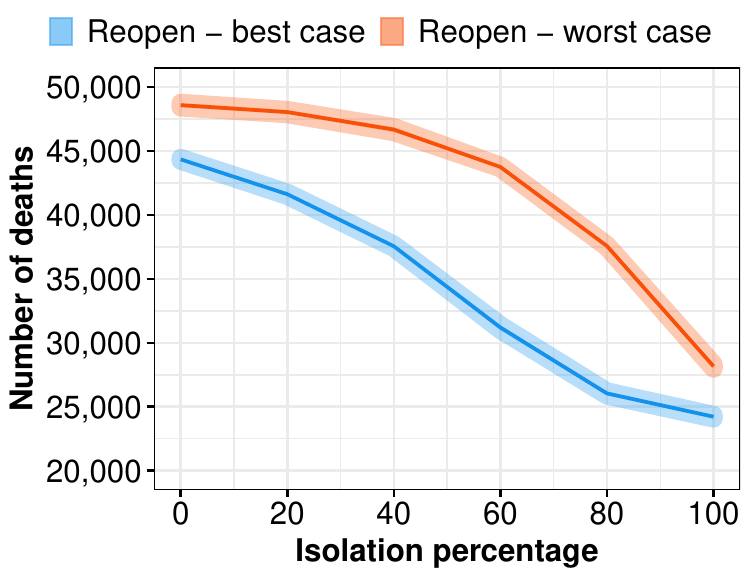}
	\caption{Student's families -- accumulated deaths.}
	\end{subfigure}
	\hspace{0.05\linewidth}
	\begin{subfigure}{0.46\linewidth}
	\includegraphics[]{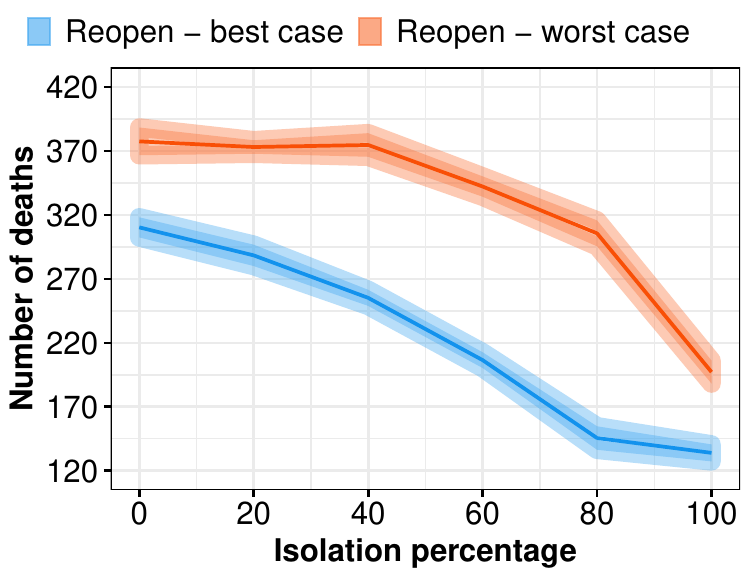}
	\caption{Teachers -- accumulated deaths.}
	\end{subfigure}

	\caption{Curves of infected people when schools reopen following São Paulo's strategy, evaluating different isolation levels.}
	\label{fig:strategy:sp:explore}
\end{figure}

\clearpage

\subsection*{Tables 1 to 4}

\begin{table}[H]
\centering
\small
\begin{tabular}{@{}lllrrrr@{}}
\toprule
Relation & Description & Type & Mean & StdDev & Min & Max \\
\midrule
Home & Number of people per home & Gamma & 3.3 & 1.7 & 1 & 10 \\
Community & Number of community relations per person & Gamma & 15.0 & 10.0 & 5 & 50 \\
Workplace & Number of people per company & Gamma & 8.0 & 7.0 & 2 & 50 \\
School -- classroom & Number of people per classroom & Gamma & 30.0 & 10.0 & 15 & 50 \\
School -- total & Number of people per school & Gamma & 500.0 & 500.0 & 300 & 4,000 \\
\bottomrule
\end{tabular}
\caption{Frequency distributions used to generate the relation network.}
\label{tb:dist:network}
\end{table}

\clearpage

\begin{table}[H]
\centering
\small
\begin{tabular}{@{}ll@{}}
\toprule
Parameter & Value \\
\midrule
$R_0$ (basic reproduction rate)   & 3.0 (at the start of the pandemic). \\
Pre-symptomatic time & 1 day. \\
Time for which people with unreported  &\\
and mild symptoms are contagious & 4 days. \\
Incubation time & Gamma, mean: 4.6 days, Stddev: 3.2, Min: 1, Max: 14. \\
Time from symptomatic to hospitalization & 5 days. \\
Days in ICU for critical patients & Gamma, mean: 15.0 days, Stddev: 15.0, Min: 1, Max: 60.\\
Days in hospital for severe patients (infirmary) & Gamma, mean: 6.5 days, Stddev: 5.5, Min: 1, Max: 60.\\
Age of the teachers & Gamma, mean: 41.2 years, Stddev: 9.9, Min: 20, Max: 70. \\
\midrule
Vaccine -- time to be effective & 14 days. \\
Vaccine -- number of people vaccinated per day & 300,000 people per day. \\
Vaccine -- immunity rate & 90\%. \\
\midrule
Infection rate -- home & $\times3.0$ the community rate considering the beginning \\
& of the pandemic (not changed by any intervention). \\
Infection rate -- workplace & $\times1.5$ the community rate. \\
Infection rate -- school & We evaluated both $\times2$ and $\times4$ the community rate. \\
Infection rate -- inter-city & Equal to the community rate. \\
Infection rate -- community & We calculate it such that the average reproduction rate  \\
& among all people matches the target reproduction rate. \\
\bottomrule
\end{tabular}
\caption{Other simulation parameters.}
\label{tb:prob:other:params}
\end{table}

\clearpage

\begin{table}[H]
\centering

\begin{tabularx}{0.8\linewidth}{@{}YYYYY@{}}
\toprule
Age & Unreported & Mild & Severe & Critical \\
\midrule
0-10  & 0.8916 & 0.1051 & 0.0024 & 0.0008 \\
10-20 & 0.9613 & 0.0384 & 0.0003 & 0.0001 \\
20-30 & 0.9447 & 0.0545 & 0.0006 & 0.0002 \\
30-40 & 0.8987 & 0.0985 & 0.0022 & 0.0006 \\
40-50 & 0.8340 & 0.1577 & 0.0064 & 0.0019 \\
50-60 & 0.7211 & 0.2513 & 0.0202 & 0.0074 \\
60-70 & 0.5560 & 0.3379 & 0.0737 & 0.0323 \\
70-80 & 0.5079 & 0.1459 & 0.2274 & 0.1188 \\
80-90 & 0.4939 & 0.2276 & 0.1854 & 0.0930 \\
90+   & 0.4931 & 0.2283 & 0.1901 & 0.0886 \\
\midrule
Mean & 0.8500 & 0.1235 & 0.0182 & 0.0084 \\
\bottomrule
\end{tabularx}
\caption{Probability of an infected person to develop specific symptoms depending on the age.}
\label{tb:prob:symptoms}
\end{table}

\clearpage

\begin{table}[H]
\centering
\begin{tabularx}{0.75\textwidth}{@{}YYY@{}}
\toprule
Age & Death (severe cases) & Death (critical cases) \\
\midrule
0-10  & 0.0109 & 0.0649 \\
10-20 & 0.0304 & 0.1405 \\
20-30 & 0.0217 & 0.1545 \\
30-40 & 0.0347 & 0.1601 \\
40-50 & 0.0574 & 0.2214 \\
50-60 & 0.1035 & 0.3086 \\
60-70 & 0.1804 & 0.4349 \\
70-80 & 0.2643 & 0.5128 \\
80-90 & 0.3944 & 0.5549 \\
90+   & 0.5230 & 0.5576 \\
\midrule
Mean & 0.1518 & 0.3742 \\
\bottomrule
\end{tabularx}
\caption{Probability of a person to die depending on their symptoms and age.}
\label{tb:prob:death}
\end{table}

\end{document}